\documentclass[%
 reprint,
nofootinbib,
 amsmath,amssymb,
 aps,
]{revtex4-1}

\usepackage{graphicx}% Include figure files
\usepackage{dcolumn}% Align table columns on decimal point
\usepackage{bm}% bold math
\usepackage{amsmath}    % For mathematical symbols and environments
\usepackage{graphicx}   % For including images
\usepackage{hyperref}   % For hyperlinks
\usepackage{cleveref}   % For enhanced cross-referencing
\usepackage{siunitx}
\usepackage{physics}
\usepackage{aas_macros}
\usepackage{natbib}

\usepackage[table,svgnames,dvipsnames]{xcolor}
\usepackage[mathlines]{lineno}
\usepackage[normalem]{ulem}
\newcommand{\kmsMpc}{\,\mathrm{km\,s^{-1}\,Mpc^{-1}}}
\crefname{figure}{Fig.}{Figs.}     % Singular and plural for figures
\Crefname{figure}{Fig.}{Figs.}     % Capitalized version for beginning of sentences

\crefname{section}{Section}{Section}

\begin{document}

\preprint{APS/123-QED}

\title{Determining the Hubble Constant through Cross-Correlation of Galaxies and Gravitational Waves}% Force line breaks with \\
%\thanks{A footnote to the article title}%
 
\author{Jiaming Pan}
\email{jiamingp@umich.edu}
\author{Dragan Huterer}
\author{Camille Avestruz}
\author{Damon H. T. Cheung}
\author{Emery Trott}
\affiliation{
Department of Physics and Leinweber Institute for Theoretical Physics, University of Michigan, 450 Church St, Ann Arbor, MI 48109
}
\author{Neal Dalal}
\affiliation{Perimeter Institute for Theoretical Physics
31 Caroline St. N, Waterloo, Ontario N2L 2Y5, Canada}
\author{Donghui Jeong}
\affiliation{Department of Astronomy and Astrophysics and Institute for Gravitation and the Cosmos, The Pennsylvania State University, University Park, PA 16802, USA}
\affiliation{School of Physics, Korea Institute for Advanced Study, Seoul 02455, Korea}
%      \author{Donghui Jeong}
%             \affiliation{Department of Astronomy and Astrophysics and Institute for Gravitation and the Cosmos,
% The Pennsylvania State University, University Park, PA, 16802, USA}
% \altaffiliation[Also at]{School of Physics, Korea Institute for Advanced Study (KIAS), 85 Hoegiro, Dongdaemun-gu, Seoul, 02455, Republic of Korea}
%       \author{Neal Dalal}
%        \affiliation{Perimeter Institute for Theoretical Physics, 31 Caroline Street N., Waterloo, Ontario, N2L 2Y5, Canada}

%\altaffiliation[Also at]{Physics Department, XYZ University.}

%Lines break  [Also at]automatically or can be forced with \\
%
 % \affiliation[Also at]{Department of Physics, University of Michigan, Ann Arbor, MI, USA}
%  \email{Second.Author@institution.edu}
% \affiliation{%
%  Authors' institution and/or address\\
%  This line break forced with \textbackslash\textbackslash
% }%

% \collaboration{MUSO Collaboration}%\noaffiliation

% \author{Charlie Author}
%  \homepage{http://www.Second.institution.edu/~Charlie.Author}
% \affiliation{
%  Second institution and/or address\\
%  This line break forced% with \\
% }%
% \affiliation{
%  Third institution, the second for Charlie Author
% }%
% \author{Delta Author}
% \affiliation{%
%  Authors' institution and/or address\\
%  This line break forced with \textbackslash\textbackslash
% }%

% \collaboration{CLEO Collaboration}%\noaffiliation

\date{\today}% It is always \today, today,
             %  but any date may be explicitly specified

\begin{abstract}
Gravitational wave (GW) standard sirens have the potential to measure the Hubble constant $H_0$ in the local universe independently of the distance ladder, and thus offer unique new insights into the Hubble tension. A key challenge with standard sirens is detecting their electromagnetic counterparts, and therefore assigning redshifts to the measured distances. One promising way to proceed is to utilize GW `dark sirens' -- events without an identified electromagnetic counterpart -- and cross-correlate their angular distribution with that of galaxies. We present a quantitative study of how precisely the Hubble constant can be measured using tomographic cross-correlation between galaxies and GW sources. Overall, we find that the constraints on $H_0$ will be limited by the quality and quantity of GW data. We find that percent-level constraints on $H_0$ will primarily depend on achieving small distance uncertainties ($\sigma_{d_L}=0.1\,d_L$), obtaining a large number of GW dark sirens ($\gtrsim$$5{,}000$), and accurate sky localization in the tomographic analysis. 
\end{abstract}

%\keywords{Suggested keywords}%Use showkeys class option if keyword
                              %display desired

\maketitle

%\tableofcontents

\section{\label{sec:intro}Introduction}

Precise measurements of cosmological parameters from an impressive array of probes have significantly refined our understanding of the Universe’s expansion. Despite these successes, accurately determining the Hubble constant ($H_0$) remains a central task in cosmology. Measurements based on nearby type Ia supernovae "anchored" by absolute distances coming from Cepheid variables yield $H_0$ with nearly a percent-level uncertainty \cite{Riess2021}. Conversely, multiple cosmological measurements at higher redshifts, including analyses of the cosmic microwave background ( CMB), big bang nucleosynthesis, and baryon acoustic oscillations, provide a much lower value of $H_0$, with measurements that are mutually consistent \cite{eBOSS:2020yzd,Planck:2018vyg,DESI:2025zgx,ACT:2025fju} and robust with respect to frequently studied extensions of the cosmological models beyond $\Lambda$CDM \cite{ACT:2025tim,DESI:2025zgx}. This disagreement between local and global measurements constitutes the so-called “Hubble tension”, whose statistical significance reaches or exceeds $5\sigma$ \cite{Freedman:2017yms,Verde:2019ivm,Knox:2019rjx,DiValentino:2021izs}.  While the global measurements come from a variety of cosmological probes with different physical dependencies on $H_0$, the local measurements are uniquely based on nearby distance indicators. This implies an urgent need for alternate measurements at low redshift that could confirm or refute the higher value of $H_0$ presently inferred from the distance ladder. Given that the difference between the distance ladder and global measurements is about $6\kmsMpc$ or $\sim$10\% of the Hubble constant, it has become clear that a percent-level independent measurement of $H_0$ will be required to decisively weigh in on the Hubble tension. 

Gravitational wave (GW) standard sirens have recently emerged as a promising alternative for directly inferring cosmological parameters \cite{Schutz:1986gp,Holz:2005df}. GW events, primarily from binary black hole (BBH) mergers \cite{KAGRA:2021vkt}, provide direct measurements of luminosity distances inferred from gravitational waveform. After assigning redshifts to GW observations by observing their respective electromagnetic (EM) counterparts, the value of $H_0$ can be inferred \cite{Nissanke:2013fka,LIGOScientific:2017adf}. However, most GW events do not have a corresponding detection of an EM counterpart; these events are commonly referred to as ``dark sirens" \cite{Finn:1994cg,MacLeod:2007jd,Messenger:2011gi,DelPozzo:2012zz}. It is nevertheless possible in principle to extract $H_0$ from dark sirens by statistically matching the location of each GW event to a number of galaxies that reside in the corresponding localization region on the sky.
% \cite{Schutz:1986gp,Nair:2018ign,Chen:2017rfc,Gray:2019ksv,Feeney:2018mkj,Gair:2022zsa}, though this method is not without challenges of its own \cite{Trott:2021fnx}. 
Conventionally, such dark siren analyses employ a Bayesian framework to integrate likelihoods from both EM observations and GW events \cite{Chen2018,DES:2019ccw,DES:2020nay,Palmese:2021mjm,Alfradique:2023giv,Gair:2022zsa}. This approach relies on statistically matching up potential host galaxies for each GW event, but may be biased unless an extremely precise localization of dark sirens is available \cite{Trott:2021fnx,Hanselman2025,Alfradique:2025tbj,Zazzera:2024agl}.

Another approach is the so-called ``spectral siren’’ method, which infers redshifts distribution from features in the compact binary mass spectrum rather than from host galaxies \cite{Chernoff:1993th,Taylor:2012db,Farr:2019twy,You:2020wju,Mastrogiovanni:2021wsd,Ezquiaga:2022zkx,Mali:2024wpq}. While promising in principle, this method can become strongly degenerate with $H_0$ if the those spectral features in the source evolve with redshift. In such cases, the nuisance evolution can mimic the $H_0$ response, inflating $\sigma(H_0)$ and erasing the $H_0$ information in the fully degenerate limit. We illustrate this behavior with a simple toy model in Appendix~\ref{app:spectral_siren}. This limitation motivates the alternative we pursue in this paper.

The most promising known alternative to standard candles is to use the standard sirens to constrain the Hubble constant. While this is much more straightforward for the bright sirens than for the dark ones, a good way to proceed in the latter case is to cross-correlate these GW events with a population of galaxies. This has the major advantage that it does not require matching up GW events with their potential host galaxies (and letting the galaxies ``vote", as in the standard dark-siren approach described above). This approach exploits the fact that both populations, GW and galaxies, trace the same underlying matter density field, so they are guaranteed to correlate, and this correlation is expected to be simple on sufficiently large scales where biases of tracers of the large-scale structure are linear. Therefore 
spatial cross-correlation between the GW events and galaxies can directly measure $H_0$ \cite{Ferri:2024amc,Oguri:2016dgk,Bera:2020jhx,Mukherjee:2022afz,Ghosh:2023ksl,Diaz:2021pem,Pedrotti:2025tfg,Beltrame:2024cve,Scelfo:2018sny}. Additionally, systematics of how each GW event corresponds to EM galaxies in the standard dark-siren analysis are treated much more simply in the cross-correlation approach than in the standard per-event analysis which employs various astrophysical efficiency and correction factors (e.g., \cite{LIGOScientific:2019zcs}). Therefore, the cross-correlation approach has the potential to simplify and make more robust the determination of $H_0$ from dark sirens. 

In this paper we systematically explore the efficacy of the galaxy–GW cross-correlation method to constrain \(H_0\). Our analysis investigates the impact on the forecasted $H_0$ precision of a variety of observational and modeling assumptions, including the choice of tomographic binning, the uncertainty in GW distance measurements, the number density of dark siren events, their redshift range and distribution, and the angular localization errors—on the statistical errors. A detailed discussion of these assumptions and their implementation is provided in Section~\ref{sec:method}, while we present our results on the impact of these assumptions in Section~\ref{sec:results}. In Section~\ref{sec:conclusion} we conclude and discuss the prospects for achieving a precise, percent‐level measurement of \(H_0\) using the galaxy–GW cross-correlation technique.

\section{\label{sec:method}Methodology}
\subsection{Cross-correlation between galaxies and GW events}

We consider two distinct populations of objects:
\begin{enumerate}
    \item \textbf{gal:} A sample of photometric galaxies, where each galaxy’s redshift distribution is given by $p^{\rm gal}_i(z)$.
    \item \textbf{gw:} Gravitational-wave events, where each event’s $d_L$ distribution that is given by $p^{\rm gw}_i(d_L)$.
\end{enumerate}
We account for the observational uncertainties in these measurements by modeling the respective redshift and distance distributions and, for simplicity, treating them as Gaussians.  For the galaxy sample, the redshift probability distribution for the \(i\)th galaxy is modeled as a Gaussian
\begin{equation}
    p^{\rm gal}_i(z) = \frac{1}{\sigma_{z_i}\sqrt{2\pi}} \exp\!\left[-\frac{1}{2}\left(\frac{z_i - z}{\sigma_{z_i}}\right)^2\right],
    \label{eq:pgal}
\end{equation}
where the uncertainty associated with the redshift measurement of the \(i\)th galaxy is given by \(\sigma_{z_i}=0.013(1+z_i)^3\), following the relation adopted by \cite{DES:2019ccw}. For the GW sample, the luminosity distance \(d_L\) for the \(i\)th event is modeled also by the Gaussian:
\begin{equation}
    p^{\rm gw}_i(d_L) = \frac{1}{\sigma_{d_{L,i}}\sqrt{2\pi}} \exp\!\left[-\frac{1}{2}\left(\frac{d_L(z,H_0) - d_{L,i}}{\sigma_{d_{L,i}}}\right)^2\right].
        \label{eq:pgw}
\end{equation}
We assume a constant relative uncertainty with the fiducial value \(\sigma_{d_{L,i}} = 0.2\,d_{L,i}\) \cite{Chen2018}. We have also checked the case where the distance error scales inversely with the signal-to-noise ratio (see Appendix~\ref{app:GW_Dis} for details).

To obtain number density distributions for the two samples, we sum the probability distributions for all sources and normalize by the total number of objects, and we get:
\begin{equation}
n^{\rm gal}(z) \propto \frac{1}{N_{\rm gal}} \sum_{i=1}^{N_{\rm gal}} p^{\rm gal}_i(z)
\overset{H_0}{\implies} n^{\rm gal}(r)
\label{eq:galaxy_n}
\end{equation}

\begin{equation}
n^{\rm gw}(d_L) \propto \frac{1}{N_{\rm gw}} \sum_{j=1}^{N_{\rm gw}} p^{\rm gw}_j(d_L)\overset{d_L/(1+z)}{\implies} n^{\rm gw}(r),
\label{eq:gw_n}
\end{equation}
where the texts over the implication signs indicate that (in Eq.~\ref{eq:galaxy_n}) we convert the galaxy redshift distribution to comoving distance by assuming a value for the Hubble constant (which is varied in the analysis), while (in Eq.~\ref{eq:gw_n}) for GW sources we  enact a simple conversion from the measured luminosity distance to comoving distance. This way, both galaxies and GW sources are represented as functions in comoving-distance space, with the important built-in dependence on the Hubble constant.

To incorporate tomography, we further bin these comoving number density distributions into a number of radial bins. For the galaxy sample, the normalized number density distribution in the \(i\)th redshift bin is defined as

\begin{equation}
n^{\rm gal}_i(r) = \frac{1}{N^{\rm gal}}\frac{dN^{\rm gal}}{dr}\,W^{\rm gal}_i(r),
\label{eq:one}
\end{equation}
where \(N^{\rm gal}\) is the total number of galaxies, \(r\) is the comoving distance, and \(W^{\rm gal}_i(r)\) is the window function that selects the contributions for each bin of galaxy sample. The differential count \(\frac{dN^{\rm gal}}{dr}\) is obtained by summing the $p^{\rm gal}_i(z)$ for all galaxies and converting to comoving distance:
\begin{equation}
\frac{dN^{\rm gal}}{dr} = \sum_{i=1}^{N_{\rm gal}} p^{\rm gal}_i(z(r)) \frac{dz}{dr}
\label{eq:dN_dr_gal}
\end{equation}
where \(p^{\rm gal}_i(z)\) is the redshift probability distribution for the \(i\)th galaxy.

A similar binning procedure applies to the GW sample. In that case, the number density distribution in the \(i\)th redshift bin is given by
\begin{equation}
n^{\rm gw}_i(r) = \frac{1}{N^{\rm gw}}\frac{dN^{\rm gw}}{dr}\,W^{\rm gw}_i(r),
\label{eq:gw_nj}
\end{equation}
with \(N^{\rm gw}\) defined as the total number of GW events. A similar procedure applies to the gravitational-wave (GW) sample, and the differential count \(\frac{dN^{\rm gw}}{dr}\) is given by:
\begin{equation}
\frac{dN^{\rm gw}}{dr} = \sum_{i=1}^{N_{\rm gw}} p^{\rm gw}_i(d_L(r))\frac{dd_L}{dr}
\end{equation}
where $p^{\rm gw}_i(d_L)$ is the luminosity distance probability distribution for the $i$th GW event.

For a bin bounded by comoving-distance values \(i_a\) and \(i_b\), the window function for tracer X (either galaxies or GW events) is specified by
\begin{equation}
W^X_i(r) = \frac{I(i_a,i_b,r)}{I(0,\infty,r)},
\label{eq:window}
\end{equation}
with
\begin{equation}
I(a,b,r) = \frac{1}{\sqrt{2\pi}\,\sigma_r} \int_{a}^{b} dr_{\rm obs} \, \exp\!\left[-\frac{(r_{\rm obs}-r)^2}{2\sigma_{r}^2}\right].
\label{eq:I}
\end{equation}
The normalization factor \(I(0,\infty,r)\) ensures that the window function integrates to unity over the entire range. In order to maximize the amplitude of the angular power spectrum when \(H_0\) assumes its true value, identical window function boundaries are adopted for both the galaxy and the GW event distributions in comoving distance. 

The cross-correlation of these two normalized number density distributions allows us to extract robust cosmological information, and specifically to constrain the Hubble constant \(H_0\)\footnote{Throughout our analysis, we assume General Relativity as the underlying theory of gravity (for alternative cosmologies, see, e.g., \cite{Belgacem:2018lbp,Nishizawa:2017nef}).}. The radial positions of galaxies are measured in redshift space, while gravitational-wave measurements report their luminosity distance; the two sets of sources are matched radially through the Hubble constant $H_0$ that links redshift to luminosity distance. This makes the tomographic cross-correlation sensitive to $H_0$, as only the true value of $H_0$ leads to the match of radial positions of GW sources and galaxies and hence produces the correct power in the tomographic cross-correlations. We further discuss the $H_0$ information from cross-correlation in Appendix~\ref{app:H0_constraint}.

The angular power spectrum is given by
\begin{equation}
P^{(\rm gal,gw)}_{ij}(\ell) = 4\pi \int_0^\infty \frac{dk}{k} \ \alpha^{\rm gal}_{i}(\ell,k)\alpha^{\rm gw}_{j}(\ell, k, H_0)\Delta_{ij}^2(k),
\label{eq:two}
\end{equation}
where the dimensionless power spectrum is separable, at least in linear theory, as
\begin{equation}
\Delta^2_{ij}(k) = \Delta^2(k,z=0) D(z_i) D(z_j), 
\label{eq:DeltaPS}
\end{equation} 
where $D(z)$ is the linear growth factor. In Eq.~(\ref{eq:two}), we make use of the projections of three-dimensional functions onto the harmonic and Fourier space; they are defined as 
\begin{widetext}
\begin{equation}
\begin{array}{rcl}
\displaystyle \alpha^{\rm gal}_i (\ell, k) &=& \displaystyle \int_0^\infty d r \, D(r, H_0) n^{\rm gal}_i (r) j_\ell(kr) b_i^{\rm gal}(z) \vphantom{\bigg(} \\
&=& \displaystyle \int_0^\infty d r \, D(r, H_0) n^{\rm gal}_i (r) j_\ell(kr)  \frac{b_i^{\rm gal}}{D(z_i)} \vphantom{\bigg(} \\
&=& \displaystyle b^{\rm gal}_i \int_0^\infty dr \, n^{\rm gal}_i (r) j_\ell(kr) \vphantom{\bigg(},
\end{array}
\label{eq:alpha_gal}
\end{equation}
\end{widetext}
where we have assumed model for the time-dependent bias where it scales as $1/D(z)$ based on \cite{DESI:2016fyo}, so that the linear growth ($D(z)$) terms cancel. Equivalently, $\alpha^{\rm gw}_j$ is defined as
\begin{equation}
\alpha^{\rm gw}_j = b_j^{\rm gw}\ \int_0^\infty dr\, n^{\rm gw}_j(r, H_0)\, j_\ell(kr).
\end{equation} 
The observed angular power spectrum, including the contribution from shot noise, is then given by
\begin{equation}
C_{ij}^{\rm gal,gw}(\ell) = P_{ij}^{\rm gal,gw}(\ell)+  P_{\ell,\mathrm{shot}}^{\,\mathrm{gal}_i\mathrm{gw}_j},
\label{eq:tildeCij}
\end{equation}
where \(i\) and \(j\) indicate the corresponding radial bins. The derivation and definitions of the shot noise terms, including the effects of angular smoothing, are provided in Appendix~\ref{app:shot_noise_smoothed}. This definition of \(C_{ij}^{\rm gal,gw}(\ell)\) ensures that both the cosmological signal and the contribution from shot noise are accounted for, and equivalently applies to \(C_{ij}^{\mathrm{gal,gal}}(\ell)\) and \(C_{ij}^{\mathrm{gw,gw}}(\ell)\).

In our fiducial test, GW events are sampled from the same redshift distribution as the galaxies. However, this does not imply that every GW event’s host galaxy is included in the galaxy catalog, since the two samples have different window functions. We have also tested more realistic GW distributions (see \cref{sec:results}) and find consistent conclusions. 

We wish to utilize \textit{only} the information on $H_0$ from angular clustering of GW and galaxies, and explicitly exclude information from the three-dimensional power spectrum. We do so in order to isolate the new information that the GW-galaxy tomography provides on top of that provided by various other measurements of three-dimensional clustering in cosmology. To remove constraints on $H_0$ arising from the dimensionless matter power spectrum $\Delta_{ij}^2(k)$, we fix $\Delta^2(k, 0)$ to the fiducial cosmology and do not vary it in our analysis. As shown in Eq.~(\ref{eq:alpha_gal}), the growth function cancels out so that the whole time-dependent power spectrum, $\Delta^2(k, z)$ can be considered fixed. As a result, because we work in flat \(\Lambda\)CDM, the luminosity distance \(d_L(z)\) and redshift relation $z$ is fully specified by just two parameters, \(\{H_0\), \(\Omega_m\}\).

%Figure \ref{fig:Pk} confirms this.

\subsection{Angular smoothing}
\label{sec:smoothing}

In our analysis, observational uncertainties are incorporated by smoothing the underlying distributions with Gaussian kernels. 
% Because convolution in real space becomes multiplication in Fourier (or spherical harmonic) space, this approach naturally introduces a window function that account for finite resolution. 
While radial smoothing has already been applied to the number density \( n^{\rm gal}_i(r) \) and \( n^{\rm gw}_j(r) \), we now include angular smoothing to consider uncertainties in the positions (see Appendix~\ref{app:radial_smoothing} for a derivation).

% \subsubsection{Angular Smoothing}
Angular uncertainties are incorporated by convolving the sky map with a Gaussian beam,
\begin{equation}
    B(\theta) = \frac{1}{2\pi \sigma_\theta^2}\,\exp\left(-\frac{\theta^2}{2\sigma_\theta^2}\right).
    \label{eq:Btheta_short}
\end{equation}
In spherical harmonic space, this convolution is equivalent to multiplying the original coefficients by the angular window function,
\begin{equation}
    W_\ell = \exp\left[-\frac{1}{2}\ell(\ell+1)\sigma_\theta^2\right].\quad
    \label{eq:Wl_short}
\end{equation}

% \subsubsection{Application to the Angular Cross-Correlation}
The effect of angular smoothing is directly incorporated into the cross-correlation analysis. 
For example, the \(\alpha\) function for GW events is modified to include the angular window function as follows 
\begin{equation}
    \alpha^{\rm gw}_{j} (\ell, k, H_0) = b_j^{\rm gw}\,W^{\rm gw}_\ell \int_0^\infty dr\, n^{\rm gw}_j(r, H_0)\, j_\ell(kr).
    \label{eq:alpha_eff_short}
\end{equation}
Accordingly, the cross-correlation with shot noise between galaxy and GW becomes
\begin{widetext}
\begin{equation}
    C^{(\rm gal,gw)}_{ij}(\ell) = 4\pi\, b_i^{\rm gal}\, b_j^{\rm gw}\, W^{\rm gal}_\ell W^{\rm gw}_\ell \int_0^\infty \frac{dk}{k}\, \left[\int_0^\infty dr\, n^{\rm gal}_i(r)\, j_\ell(kr)\right]\left[\int_0^\infty dr\, n^{\rm gw}_j(r, H_0)\, j_\ell(kr)\right] \Delta_{ij}^2(k)+  P_{\ell,\mathrm{shot}}^{\,\mathrm{gal}_i\mathrm{gw}_j}.
    \label{eq:crosscorr_short}
\end{equation}
\end{widetext}

Eq.~\eqref{eq:crosscorr_short} is the expression for the angular power spectrum used in our analysis. A derivation of these expressions, including the treatment of angular smoothing, is provided in Appendix~\ref{app:angular_smoothing}. The auto-correlation expressions $C^{(\rm gal,gal)}_{ij}(\ell)$ and $C^{(\rm gw,gw)}_{ij}(\ell)$ follow the same form as Eq.~\eqref{eq:crosscorr_short}, with the appropriate substitutions for the tracer populations and shot noise terms.

\begin{figure}[t]

\includegraphics[width=1.16\columnwidth]{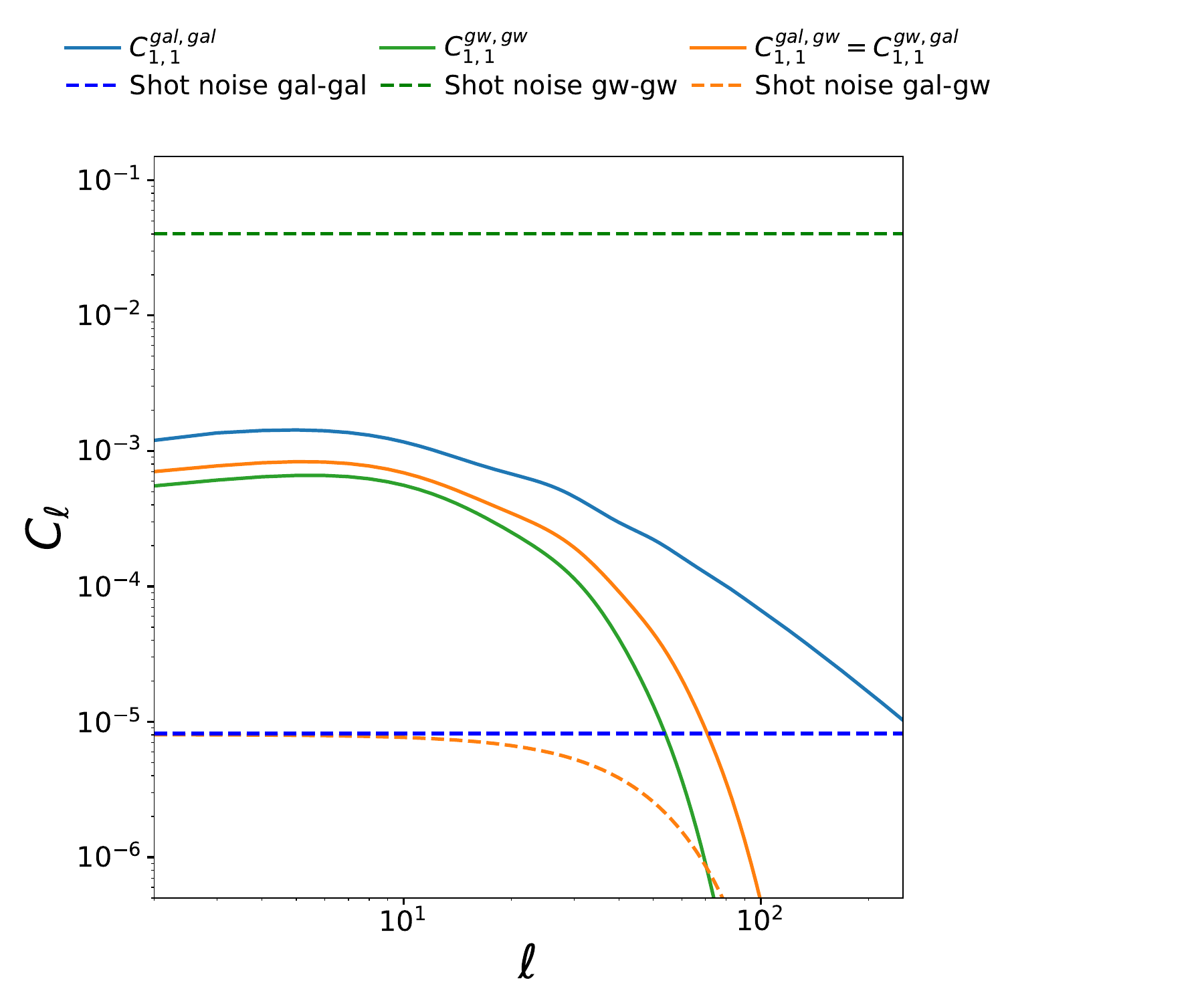}
\caption{Angular power spectra $C_{\ell}$ for galaxies and gravitational-wave events in the first bin with $z\in[0.1,0.15)$, assuming fifteen radial bins in total. The horizontal lines indicate the corresponding shot noise for each population. 
We show the auto-correlation power spectra of galaxies and GWs, as well as their cross-correlation power spectrum.
Note that the shot noise in the GW auto-correlation is particularly large due to the relatively small number density of GW sources (here, 
we assume 1{,}000 GW events over a 10{,}000~deg$^2$ area). The shot noise cross terms are suppressed at high $\ell$ due to large angular localization error of GW sources.}
\label{fig:Cls_1}
\end{figure}

Figure \ref{fig:Cls_1} shows the angular power spectrum of galaxies and gravitational-wave (GW) events for the first bin (assuming a total of fifteen bins). The shot noise in the GW, galaxies, and their cross, are shown separately as dashed lines. Note that the shot noise of GW sources dominates over the signal (the angular power spectrum) and over the galaxy and cross noise, reflecting the much lower number density of GW sources compared to galaxies. At high multipoles ($\ell$), the cross-terms of shot noise are suppressed, owing to the large angular localization uncertainty of GW events.

\subsection{\label{sec:citeref}Setup for testing the cross-correlation method}

We adopt an analytical form to model the galaxy redshift distribution inspired by \cite{Lima2008}. Specifically, the galaxy redshift distribution is modeled as
\begin{equation}
n_z(z) \propto z^2 \exp\!\left[-\left(\frac{z-0.2}{z_0}\right)^2\right],
\label{eq:nz_galaxies}
\end{equation}
where \(z_0\) is a parameter controlling the width of the distribution (here, we choose \(z_0=0.25\)). This expression captures both the \(z^2\) scaling—reflecting the growth of the cosmic volume element at low redshift—and an exponential cutoff at higher redshift. We assume that the GW events follow the same radial distribution as the galaxies, and we select GW from the same radial distribution as the galaxies given in Eq.~(\ref{eq:nz_galaxies}).

To complement this analytical model of the distribution of galaxies (as well as GW events which are selected from the galaxy sample), we also performed tests with the galaxy redshift distribution using simulated galaxy data from the MICECAT catalog \cite{Fosalba:2013wxa,Crocce:2013vda,Fosalba:2013mra}. We find that our overall conclusions remain largely unchanged when using simulated data from MICECAT. We eventually chose the analytical form to produce the results because it allows us to easily test different redshift distribution scenarios. Additionally, we conducted our analysis with a GW population sampled from the GWTC-3–inferred model of LIGO, Virgo, and KAGRA (LVK) in Appendix~\ref{app:GW_Dis}, and found that this does not change our results.

We choose to fix $\ell_{\max} = 250$ for our analysis. We have verified that for our fiducial setup, the constraint on $H_0$ converges by this choice (see \cref{sec:results} for details). Finally, to improve computational efficiency, we compute the angular power spectrum using a bin width of $\Delta\ell=10$ in $\ell$-space; we have checked that this approximation does not lead to appreciable information loss relative to using all multipoles.

\begin{figure*}[t]
\centering
\includegraphics[width=1.0\linewidth]{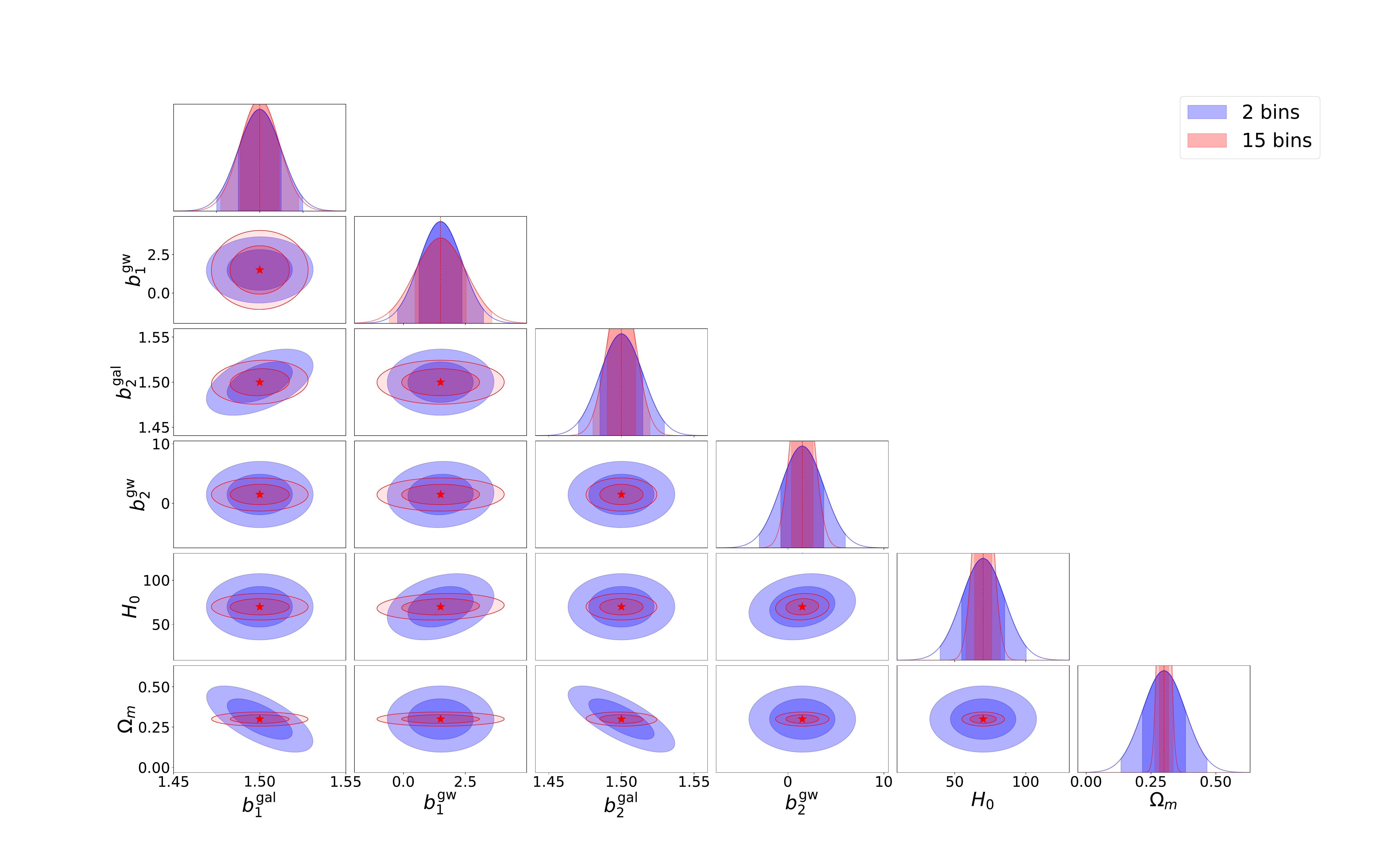}
\caption{Forecasted constraints on key cosmological parameters from the cross-correlation between galaxies and GW dark sirens. The plot includes results for both bin 2 and bin 15 cases, and assuming GW event number $N_{\rm GW}=1000$. The red stars denote the true values of the parameters. The shaded regions represent the 1$\sigma$ and 2$\sigma$ (68.3\% and 95.4\% credible levels in 2D) intervals. Increased number of radial bins breaks the parameter degeneracies, thereby providing more stringent constraints on the Hubble constant $H_0$. Note that the uncertainty on the first bias parameter, $b^{\rm gal}_1$, is comparable in the 15-bin and 2-bin cases because there are fewer objects in that bin in the former case, but the constraints on the Hubble constant nevertheless improve with more tomographic bins as expected.}
\label{fig:cross_fisher}
\end{figure*}

\subsection{Fisher matrix forecasts}

In our analysis, we employ the Fisher matrix forecast to quantify the statistical errors and covariances of estimated parameters. The Fisher matrix is computed as \cite{McQuinn:2013ib}
\begin{equation}
    F_{ab} = f_{\text{sky}}\frac{1}{2}\sum_{\ell}(2\ell+1){\rm Tr}[A^{-1}A_{,a}A^{-1}A_{,b}],
\end{equation}
where $f_{\text{sky}}$ is the fraction of the full sky that is observed. The matrix \( A \) is given by
\begin{center}
    \( A_{ij} = \left[ \begin{array}{cc}
    C^{(\rm gal,gal)}_{ij} & C^{(\rm gal,gw)}_{ij} \\
    C^{(\rm gal,gw)}_{ij} & C^{(\rm gw,gw)}_{ij}
    \end{array} \right] \)
\end{center}
which expands for multiple bins as
\[
A_{ij} =
\left[
\begin{array}{cccc}
    C^{\text{gal,gal}}_{11} & C^{\text{gal,gw}}_{11} & \dots & \dots \\
    C^{\text{gal,gw}}_{11} & C^{\text{gw,gw}}_{11} & \dots & \dots \\
    \vdots & \vdots & \ddots & \\
    & & C^{\text{gal,gal}}_{N_{\text{bins}}N_{\text{bins}}} & C^{\text{gal,gw}}_{N_{\text{bins}}N_{\text{bins}}} \\
    \vdots & \vdots & C^{\text{gal,gw}}_{N_{\text{bins}}N_{\text{bins}}} & C^{\text{gw,gw}}_{N_{\text{bins}}N_{\text{bins}}}
\end{array}
\right]
\]

We choose the fiducial cosmology to have the matter density relative to critical of \(\Omega_m = 0.3\), and Hubble constant $H_0 = 70\, \mathrm{km/s/Mpc}$. We also assume linear bias throughout, which is consistent with our approach of using large angular (and therefore spatial) scales. We assume one bias parameter per tomographic bin, separately for GW and galaxies. Here we set the redshift-zero biases to \(b^{\rm gal}=1.5\) for the galaxy bias and \(b^{\rm gw}=1.5\) for the GW bias (and then both scale with $1/D(z)$, as explained near Eq.~(\ref{eq:alpha_gal})). Our set of parameters that are free to vary is therefore 
\begin{equation}
    p_i\in \{b^{\rm gal}_1, b^{\rm gal}_2, \dots, b^{\rm gal}_{N_{\text{bins}}}, \, b^{\rm gw}_1, b^{\rm gw}_2, \dots, b^{\rm gw}_{N_{\text{bins}}}, \, H_0, \, \Omega_m\},
    \label{eq:pars}
\end{equation}
where $N_{\text{bins}}$ is the number of tomographic bins.

The uncertainty on the Hubble constant is defined as $\sigma(H_0) \simeq \sqrt{ \left[ \mathbf{F}^{-1} \right]_{H_0 H_0} }$
, where \(\left[ \mathbf{F}^{-1} \right]_{H_0 H_0}\) refers to the diagonal element of the inverse Fisher matrix corresponding to the Hubble constant. Interestingly we find that the marginalized $H_0$ error (to be reported below) is only modestly larger than that in the fictitious case when all of the other parameters are fixed; this indicates that the cross-correlations are able to break degeneracy between the bias terms and the Hubble constant. We will discuss this in further details in the section \ref{sec:results}.

Using the Fisher forecast, we systematically test several key survey and analysis specifications to assess their impact on the precision of the \( H_0 \) constraints in cross-correlation method. The inputs under investigation are:

\begin{itemize}
    \item \textbf{Number of Tomographic Bins (\(N_{\text{bins}}\))}: The fiducial analysis is performed using 15 bins. We extend the analysis up to approximately 25 bins to explore how tomography influences the constraints.
    \item \textbf{Error in GW Distance Measurements (\(\sigma_{d_L}\))}: Our fiducial relative uncertainty is \(\sigma_{d_L} = 0.2\,d_L\). An alternative scenario with a reduced error of \(\sigma_{d_L} = 0.1\,d_L\) is also tested, allowing us to evaluate the sensitivity of \( H_0 \) constraints to the luminosity distance error.
    \item \textbf{Number of Dark-Siren GW Events (\(N_{\text{GW}}\))}: We consider a range from 1,000 (fiducial) to $50,000$ events. This test allows us to quantify how increasing the number of events improves the statistical precision.
    \item \textbf{Redshift Range}: The fiducial redshift interval is set from \(z = 0.1\) to \(z = 0.7\). Alternate redshift ranges are examined to understand the effects of galaxy redshift distribution on the \(H_0\) constraints.
    \item \textbf{Error of GW Event Localization}: The fiducial GW events localization uncertainty is assumed to be 10 square degrees. We also test an alternative scenario with a more optimistic 1-square degree error. 
\end{itemize}
Note that all of these parameters are affecting the statistical errors; in this paper we do not study the impact of the systematics. Our baseline configuration assumes a survey covering \(10\,000\,\mathrm{deg}^2\) (i.e., \(f_{\mathrm{sky}} \approx 0.24\)), with \(1.2 \times 10^7\) galaxies and \(1,000\) GW events. In the following section, we show the results of these investigations.

\begin{figure*}[t]
\includegraphics[width=\linewidth]{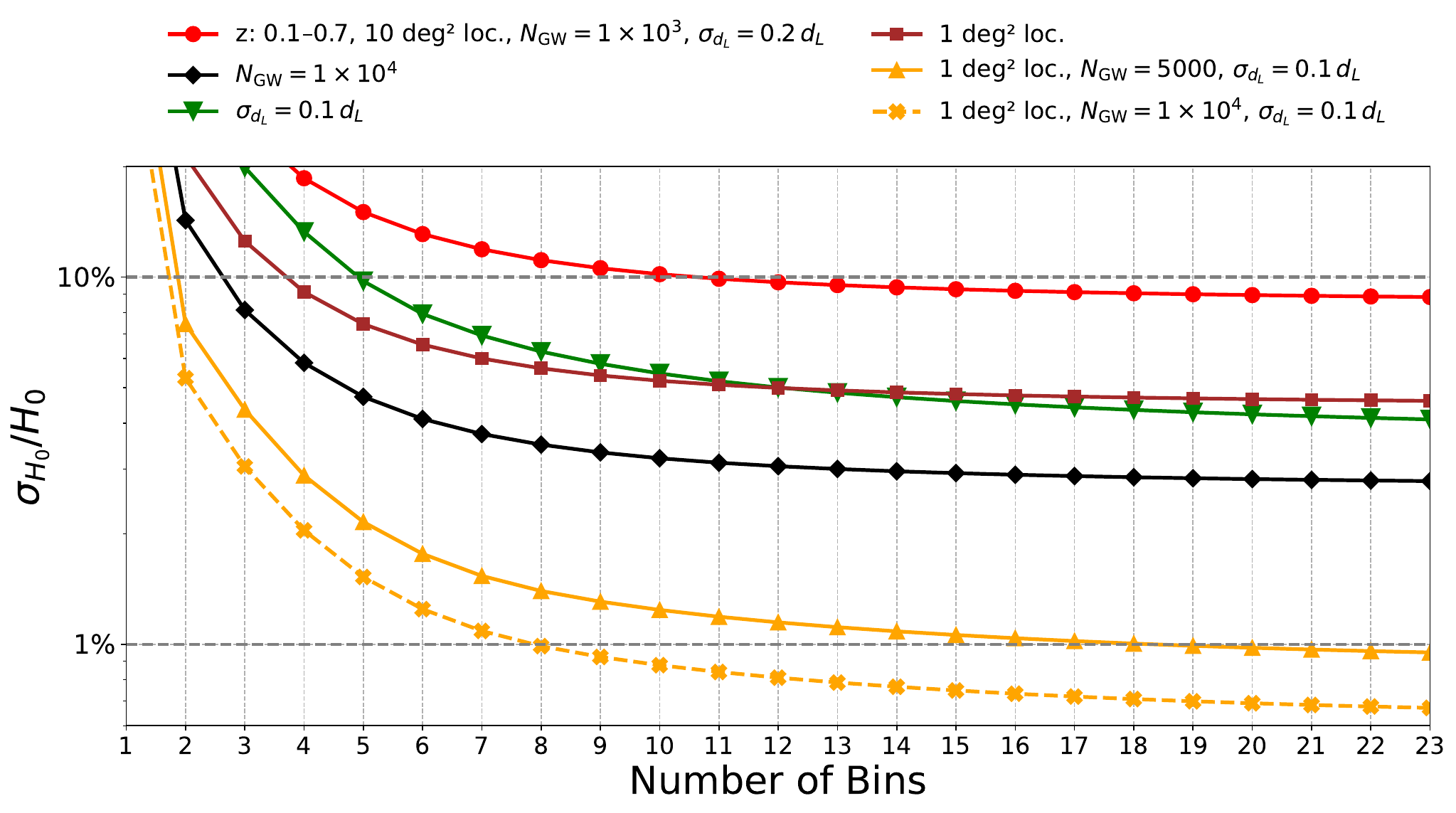}
\caption{Forecasted relative uncertainty on the Hubble constant $H_0$ as a function of the number of tomographic radial bins. 
The fiducial configuration (red) assumes $z\in[0.1,0.7]$, $10\,\mathrm{deg}^2$ localization uncertainty, $N_{\rm GW}=10^3$, and $\sigma_{d_L}=0.2\,d_L$. 
Variants shown are: improved luminosity-distance errors $\sigma_{d_L}=0.1\,d_L$ (green); reduced localization errors to $1\,\mathrm{deg}^2$ (brown); a larger GW sample size $N_{\rm GW}=10{,}000$ (black); the combined case with $1\,\mathrm{deg}^2$, $\sigma_{d_L}=0.1\,d_L$, and $N_{\rm GW}=5000$ (orange); and the same combined case but with $N_{\rm GW}=10{,}000$ (orange dashed). 
Gray dashed lines mark the $10\%$ and $1\%$ levels. This illustrates that percent-level constraints require simultaneously large $N_{\rm GW}$, precise angular localization, \textit{and} small luminosity-distance errors.}
\label{fig:H0_Combined}
\end{figure*}

\section{Results 
\label{sec:results}}

We now present results of our analysis of the GW–galaxy cross-correlation method for constraining the Hubble constant. We examine how tomographic binning improves $H_0$ constraints, explore the impact of distance errors, angular localization uncertainties, and the number of GW events, and test different redshift distributions (including the GWTC-3–inferred population). Finally, we summarize the results with an empirical scaling relation for $\sigma_{H_0}$.

\cref{fig:cross_fisher} displays the 2D confidence ellipses derived from the Fisher matrix analysis for both two and fifteen tomographic bins (the latter of which is our fiducial choice); we assume $N_{\rm GW}=1,000$ GW events, with all other settings identical to the fiducial choice. We show constraints on the four bias parameters that are featured in both of these tomographic analyses, as well as the Hubble constant and $\Omega_m$.  It is interesting that the constraints on the Hubble constant significantly improve with more tomographic bins despite the proliferation of bias parameters (whose number is, recall, is $2N_{\rm bins}$) due to the breaking of parameter degeneracies\footnote{Note that the improvement in constraining the bias terms is not uniform in each bin; the improvement with increasing number of tomographic bins in the lowest- and highest-redshift bias terms is reduced due  the decreased number of objects in the corresponding redshift ranges when $N_{\rm bins}$ increases. This is why $b^{\rm gal}_1$ doesn't improve with increased tomographic bins in \cref{fig:cross_fisher}.}. 

We have investigated the breaking of the degeneracy between the bias parameters and $H_0$ in a little more detail.  We find that breaking of the degeneracy persists when we increase the GW luminosity distance error to $\sigma_{d_L} = 0.3\,d_L$. Similarly, setting the shot noise in the cross-tracer terms (i.e., gal-gw terms) to zero does not affect significantly the fiducial constraints, indicating that the degeneracy breaking is not driven by the information from these terms. We do find, however, that the constraints dramatically worsen, and the parameter degeneracy becomes much stronger, once we exclude the off-diagonal cross terms in the angular power spectrum (that is, $C_{ij}$ when $i\neq j$). This indicates that the tomographic correlations between different radial bins are crucial in helping break the degeneracy between the bias parameters and $H_0$.

Figure~\ref{fig:H0_Combined} presents the forecasted uncertainties on \(H_0\) as a function of the number of radial bins under a variety of assumptions including the number of GW events, GW event localization errors, and GW luminosity distance uncertainties. The fiducial case is the red line, and the other curves, as detailed in the legend, are scenarios with variations relative to the fiducial assumptions.

Figure~\ref{fig:H0_Combined} shows that increasing the number of tomographic bins significantly increases the constraints on $H_0$. In the fiducial case with \(\sigma_{d_L}=0.2\,d_L\), the constraint on \(H_0\) improves rapidly and reaches convergence around $\sim$15 bins, yielding a fractional uncertainty of $\sigma(H_0)/H_0 \simeq 9\%$. Note that the fiducial forecast with $N_{\rm GW}=1{,}000$ and a fixed $\sigma_{d_L}=0.2\,d_L$ should be regarded as conservative. In realistic GW data, $\sigma_{d_L}$ varies with the signal-to-noise ratio, and a subset of nearby, high-SNR events achieves substantially better precision than $0.2\,d_L$ (e.g., \cite{Bom:2024afj}). Therefore, with $N_{\rm GW}=1{,}000$ real events, the constraints should improve over this fixed-error baseline. For a smaller error \(\sigma_{d_L}=0.1\,d_L\), a larger number of tomographic bins is required for convergence. This is because the tomographic information is exhausted when the bin width becomes comparable to, or smaller than, the typical galaxy photo-z errors.

While improvements in \(\sigma_{d_L}\) and GW localization errors (e.g., achieving a \(1\,\mathrm{deg}^2\) error) help tighten the \(H_0\) measurements, these factors alone are insufficient to reduce the uncertainty to one percent-level. For example, with $N_{\rm GW}=1{,}000$, the case with \(\sigma_{d_L}=0.1\,d_L\) yields a fractional uncertainty of \(\sigma(H_0)/H_0 \simeq 4.6\%\), while assuming \(1\,\mathrm{deg}^2\) localization gives \(\simeq 4.8\%\).

Our forecasts indicate that achieving sub-percent precision in \(H_0\) ultimately requires a substantial increase in the number of GW events. Specifically, under optimistic conditions assuming a localization error of \(1\,\mathrm{deg}^2\) and distance error \(\sigma_{d_L} = 0.1\,d_L\), we find that $5,000$ GW events or more events are necessary to reach the desired sub-percent accuracy in $H_0$. While such tight localization and distance errors will be challenging to achieve, a large number of GW events is expected with future observations using 3G GW‐detector network composed of the Einstein Telescope \cite{Abac:2025saz} and Cosmic Explorer \cite{Evans:2023euw}, making it perhaps the most ready survey-specification parameter to improve upon for driving down the uncertainty in \(H_0\). Also, we note that even with $N_{\rm GW}=50{,}000$, $\sigma_{d_L}=0.2,d_L$, and $10\,\mathrm{deg}^2$ localization (not shown in the figure), the forecasted constraint on $H_0$ saturates at $\sigma_{H_0}/H_0 \simeq 1.2\%$, i.e., it does not reach the sub-percent regime. This indicates that increasing $N_{\rm GW}$ alone is insufficient; achieving sub-percent precision will also require improved luminosity-distance accuracy and better sky localization.

Interestingly, marginalization over the galaxy and GW bias parameters does not significantly affect the $H_0$ uncertainties. The reason is that the degeneracies between galaxy biases and $H_0$ are broken with sufficient number of bins. As a result, the dominant source of error on $H_0$ is the large shot noise of GW events due to low number density.

\begin{figure}[tbp]
  \centering
  \includegraphics[width=\linewidth]{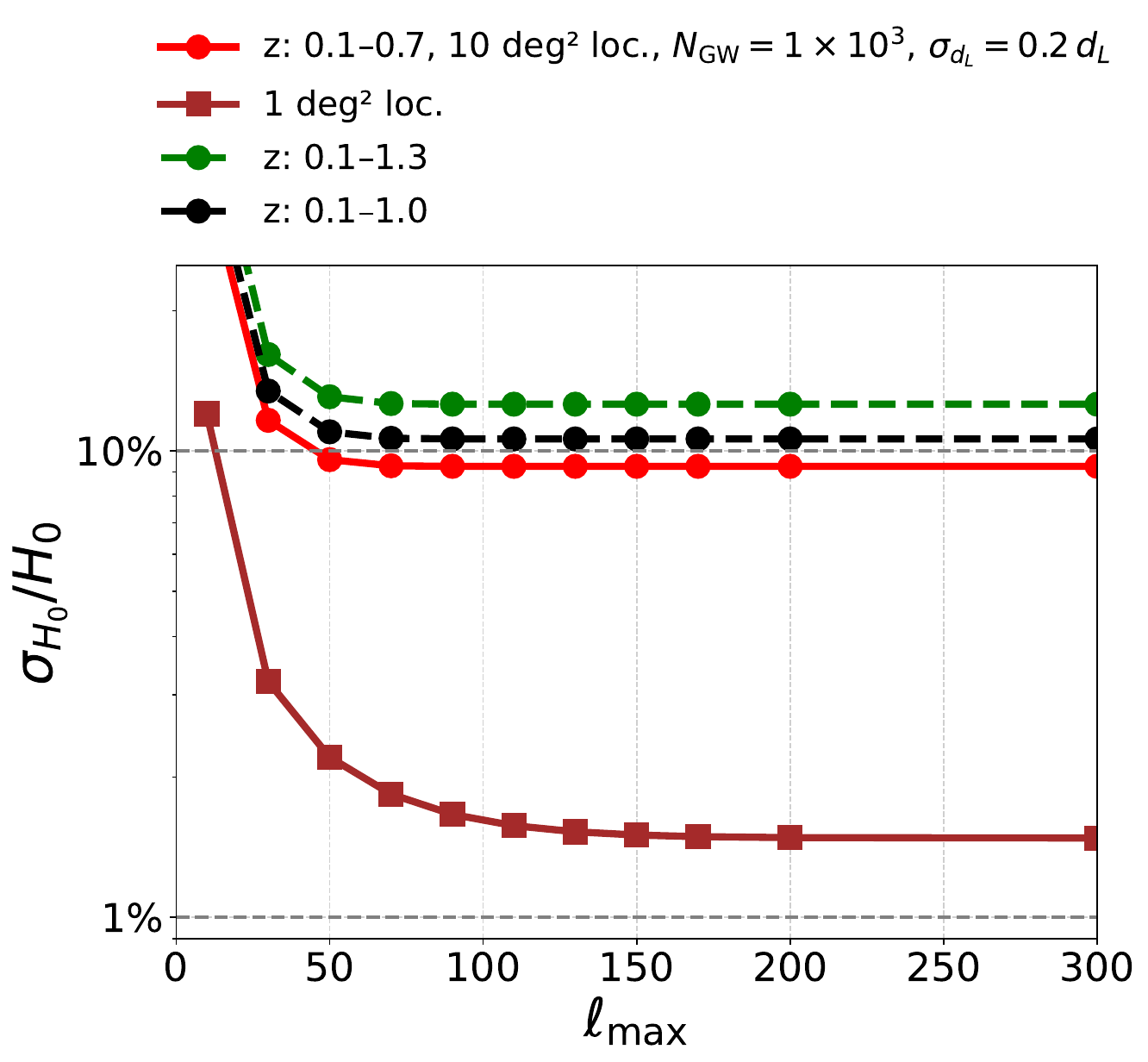}
  \caption{Forecasted fractional uncertainty on the Hubble constant $H_0$ as a function of the maximum multipole $\ell_{\max}$. We compare scenarios with different localization areas (10 deg$^2$ and 1 deg$^2$) and with broader galaxy redshift distributions ($0.1<z<0.7$, $0.1<z<1.0$, and $0.1<z<1.3$). All cases assume $N_{\rm GW}=1000$ and a distance error of $\sigma_{d_L}=0.2\,d_L$.}
  \label{fig:lmax_H0}
\end{figure}

Recall that we have adopted a fiducial choice of $\ell_{\max}=250$ for our results. \cref{fig:lmax_H0} shows the forecasted $\sigma_{H_0}$ converges by this scale for both the $10\,\mathrm{deg}^2$ and $1\,\mathrm{deg}^2$ localization scenarios, with the $10\,\mathrm{deg}^2$ case converging at smaller $\ell$ due to its lower resolution. The choice of $\ell_{\max}$ is also robust to the assumed galaxy redshift distribution. We have also compared the linear and nonlinear predictions for the angular power spectra using the halofit model \cite{2003MNRAS.341.1311S}, and determined the scale where their difference becomes statistically significant. This test shows that linear theory is valid up to $\ell\sim100$ at low redshift and up to $\ell\sim200$ at higher redshift under our fiducial assumptions. Hence, $\ell_{\max}=250$ provides an optimistic and converged cutoff, extending slightly into the mildly nonlinear regime but still retaining essentially all of the constraining power available at linear scales. We also find that broadening the galaxy redshift distribution (e.g., $0.1<z<1.0$ and $0.1<z<1.3$) does not significantly change the inferred $\sigma_{H_0}$ at fiducial $N_{\rm GW}$, angular localization uncertainty, and $\sigma_{d_L}$, and the multipole at which the forecasts converge remains $\ell\simeq100$ across all different choices of redshift distribution.

\begin{figure}[tbp]
  \centering
  \includegraphics[width=\linewidth]{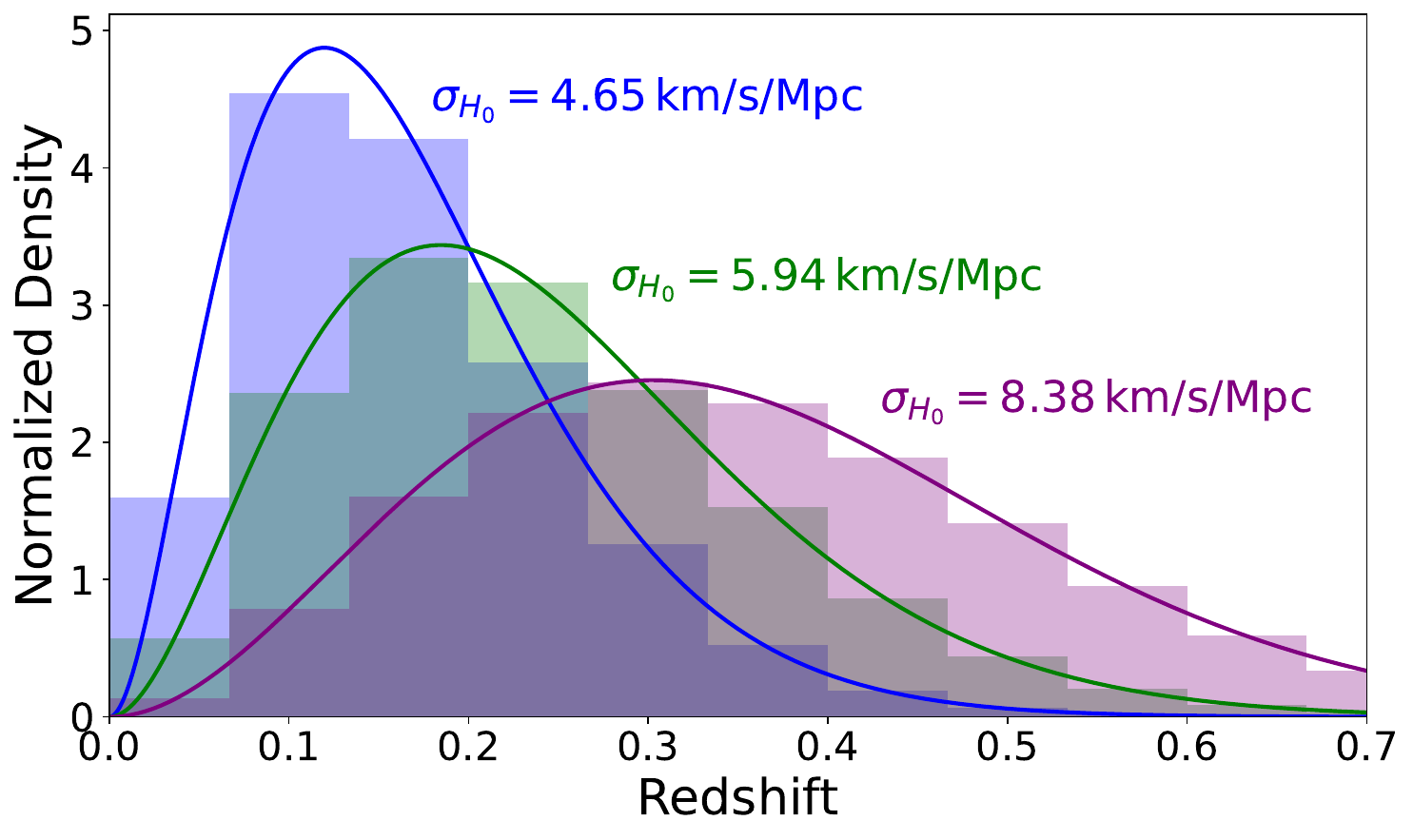}
  \caption{Dependence of forecasted errors in $H_0$ on the galaxy redshift distribution, where the GW events are drawn. We show cases with three different galaxy redshift distributions, assuming throughout a fixed galaxy number density (so that the total number of galaxies varies between the three cases). Annotations indicate the marginalized error in $H_0$ assuming in each case fifteen tomographic bins and \(N_{\rm GW}=1000\). The results indicate that low-redshift GW samples are very important for tighter $H_0$ constraints.}
  \label{fig:redshift_distribution}
\end{figure}

Additionally, we tested the impact of galaxy redshift distribution on the constraints of $H_0$  in \cref{fig:redshift_distribution}. In this test, we draw GW events from galaxy samples with different redshift ranges ($z=0.1$--$0.25$, $0.15$--$0.30$, and $0.20$--$0.35$), assuming a constant galaxy number density so that the total number of galaxies varies with the redshift selection. We find that low-$z$ GW samples provide significantly tighter $H_0$ constraints. At low redshift, the absolute distance errors are smaller and the cross-correlation signal is stronger. At higher redshift, although the survey volume and the number of galaxies increase and shot noise decreases slightly, the clustering signal weakens substantially, which reduces the overall constraining power. By testing a shot-noise–only case, we confirm that such a configuration without two-halo terms yields very weak $H_0$ constraints, demonstrating that the dominant information comes from the cross-correlation signal. Thus, low-$z$ GW samples are especially valuable for precision $H_0$ measurements. Further details on these results are provided in Appendix~\ref{app:redshift_Dis} and \cref{fig:redshift_effects}.

% Additionally, we tested the impact of galaxy redshift distribution on the constraints of $H_0$ (\cref{fig:redshift_distribution}). In this test, we draw GW events from galaxy samples with different redshift ranges ($z=0.1$--$0.25$, $0.15$--$0.30$, and $0.20$--$0.35$), assuming a constant galaxy number density so that the total number of galaxies varies with the redshift selection. We find that low-$z$ GW samples provide significantly tighter $H_0$ constraints. At low redshift, the absolute distance errors are smaller and the cross-correlation signal is stronger. At higher redshift, although the survey volume and the number of galaxies increase and shot noise decreases slightly, the clustering signal weakens substantially, which reduces the overall constraining power. Thus, low-$z$ GW samples are especially valuable for precision $H_0$ measurements. Further details on these results are provided in Appendix~\ref{app:redshift_Dis} and \cref{fig:redshift_effects}.

Moreover, because GW detectors and galaxy surveys have different selection functions, the true redshift distribution of GW events does not necessarily trace that of the galaxies. In our fiducial analysis, we drew GW events from the galaxy redshift distribution, which assume both galaxy survey and GW detector have the same selection effects. In practice, however, the observation biases for galaxy surveys and GW detectors differ. To quantify the effect, we instead sample GW events from the inferred GWTC-3~\citep{KAGRA:2021duu} population model of LIGO, Virgo and KAGRA 
 (LVK)~\cite{KAGRA:2021duu} O3 measurement. We compute each event’s luminosity distance uncertainty, $\sigma_{d_L}$, by sampling a fitted relation between signal‐to‐noise ratio (SNR) and $\sigma_{d_L}$. In this approach, more distant GW events have larger $\sigma_{d_L}$ (and the nearby GW events have smaller $\sigma_{d_L}$) compared to the case of \(\sigma_{d_L} = 0.2\,d_L\). Carrying out this test, we find that assuming different galaxy and GW distributions leads to similar $H_0$ constraints as in our fiducial case with \(\sigma_{d_L} = 0.2\,d_L\). Further implementation details and numerical results are given in Appendix~\ref{app:GW_Dis} and \cref{fig:H0_LVK}.

 To facilitate interpretation of these forecasts, we empirically determine how the $\sigma_{\rm H_0}$ scales with the GW events properties. By fitting the dependence of the $H_0$ uncertainty at $N_{\rm bins}=22$ across all tested scenarios, we find that the $H_0$ constraint approximately follows the scaling relation
 \begin{equation}
  \sigma_{H_0} \simeq 6.2\,
  \left(\frac{N}{N_0}\right)^{-0.50}
  \left(\frac{\sigma_{d_L}}{\sigma_{d_L,0}}\right)^{1.10}
  \left(\frac{\sigma_{\theta^2}}{\sigma_{\theta^2,0}}\right)^{0.28},
\end{equation}
where the prefactor has units of $\kmsMpc$; here $N_0 \equiv 10^3$, the distance error $\sigma_{d_L,0} \equiv 0.2 d_L$, and the GW localization error $\sigma_{\theta^2,0} \equiv 10~\mathrm{deg}^2$. This result indicates that the $H_0$ error decreases nearly as $N^{-1/2}$ with the number of events, is roughly proportional to the distance uncertainty. The dependence on $\sigma_{\theta^2}$ is the weakest of the three, which also agrees with the weak dependence on GW event localization uncertainty using the Bayesian approach found in \cite{Cross-Parkin:2025xwf}. This suggests that improving distance measurement precision is most important for tightening $H_0$ constraints in cross-correlation technique.

% \begin{equation}
%     % \sigma_{H_0} \propto N^{-0.50}\,\sigma_{d_L}^{1.10}\,\sigma_{\theta^2}^{0.28},
%      \sigma_{H_0} \simeq 6.21\left(\frac{N}{1000}\right)^{-0.50}\left(\frac{\sigma_{d_L}}{0.2}\right)^{1.10}\left(\frac{\sigma_{\theta^2}}{10}\right)^{0.28},
% \end{equation}

\section{Discussion and conclusions \label{sec:conclusion}}

We have conducted a quantitative study of how precisely cross-correlation between gravitational-wave events and galaxies can be used to constrain the Hubble constant. We have purposefully not adopted a configuration of any planned future GW experiment (other than in the test explained in the last paragraph of Sec.~\ref{sec:results}), but have rather varied several key parameters that characterize the quality and quantity of GW data in order to investigate how they impact the precision in $H_0$. This approach complements more specialized forecasts that aim to describe data quality expected for specific proposed surveys (e.g., \cite{Iacovelli:2022bbs}).

Our forecasts suggest that achieving sub-percent precision on \(H_0\) through the cross-correlation method may remain challenging unless a substantially larger sample of dark-siren GW events with small uncertainties in distances and very accurate localizations becomes available. Even under the highly optimistic scenario of \(N_{\rm GW}=50,000\) dark-siren events—achievable with next-generation detectors—with distance uncertainties \(\sigma_{d_L}=0.2\,d_L\) and sky localizations of \(10\,\mathrm{deg}^2\), the cross-correlation method fails to achieve sub-percent constraints on \(H_0\). We find that reaching percent-level $H_0$ constraints will require a \(1\,\mathrm{deg}^2\) localization, along with at least \(N_{\rm GW}\approx5,000\) events with \(\sigma_{d_L}=0.1\,d_L\) distance uncertainties. Gravitational-wave dark-sirens constraints on \(H_0\) that will shed light on the Hubble tension therefore critically depend on high-accuracy luminosity distance measurements and precise sky localizations for a large number of events expected from future GW surveys (see forecasted 1$\sigma$ uncertainties in Figure~\ref{fig:H0_Combined}).

We find that the $H_0$ uncertainty in cross-correlation technique follows the approximate scaling relationship $N^{-0.50}\,\sigma_{d_L}^{1.10}\,\sigma_{\theta^2}^{0.28}$
where $N$ is the number of GW events, $\sigma_{d_L}$ is the luminosity distance uncertainty, and $\sigma_{\theta^2}$ is the sky localization uncertainty in square degrees. This relation underscores the fact that reducing distance measurement errors is most impactful for improving $H_0$ constraints.

We also find that low-redshift GW and galaxy samples are particularly helpful for constraining $H_0$ (see Figure~\ref{fig:redshift_distribution}). 
In our fiducial tests we assumed that both galaxy and GW events follow the same radial distribution. 
Nevertheless, drawing GW events from the GWTC-3 population model with $\sigma_{d_L}$ determined by their signal-to-noise ratios yields $H_0$ constraints very similar to those in the fiducial case with $\sigma_{d_L} = 0.2\,d_L$ and GW events sampled from the galaxy distribution (see Figure~\ref{fig:H0_LVK}). 

Our results broadly align with recent forecasts that were posted while this paper was in preparation \cite{Ferri:2024amc,Pedrotti:2025tfg,Sala:2025wwu}. These papers assume specific configurations of planned or conceptualized surveys, rather than isolating dependence on key data-quality parameters that we adopted in this work.  We have checked that, assuming similar assumptions as these investigations, we find results that are in a good general agreement with them. In particular, \citet{Ferri:2024amc} only find sub-percent $H_0$ constraints if they assume the scenario of a next-generation GW detector with more than $50,000$ GW events. Under similar localization uncertainties and distance errors, our results are in broad agreement. Similarly, \citet{Pedrotti:2025tfg} find percent-level constraint on $H_0$ only if they assume the Einstein Telescope + Cosmic Explorer configuration with $\sim100,000$ GW  events and $1.6\times10^9$ galaxies from Euclid; we again find good agreement with them once we assume similar numbers and approximate their assumed distribution of sky-localization and distance errors.  Moreover, \citet{Pedrotti:2025tfg} allow information from the matter power spectrum (which, recall, we explicitly exclude to isolate the geometric information from cross-correlations), which further boosts their precision.

Overall then, this paper and other related work all indicate that  $H_0$ constraints sufficiently precise to decisively weigh in on the Hubble tension might be available from GW-galaxy cross-correlations, but only once we have GW dark-siren data of much higher quality and quantity than presently available. It is at present unclear \textit{when} this will be the case. Given that the Hubble tension has only become stronger over the last $\sim$5 years, it may well be that it will remain relevant for years to come, up until the next generation of GW experiments is online and operating.

Other approaches for dark sirens also exist. Recent studies have shown that the traditional Bayesian framework can lead to biases in cosmological inference~\cite{Trott:2021fnx,Hanselman2025,Alfradique:2025tbj,Zazzera:2024agl}. The “spectral siren’’ methods infer redshift information from the source-frame mass distribution, but can be affected by degeneracies with possible redshift evolution of that distribution, which may limit the precision of $H_0$ inference (Appendix~\ref{app:spectral_siren}; see also \cite{Agarwal:2024hld,Tong:2025xvd}). We leave investigation of potential biases in the cross-correlation approach and a comparison with these other methods to future work.

Finally, we note that our forecasts are conservative in that we have explicitly zeroed out any information coming from the matter power spectrum $P(k)$ and only kept the information from the cross-correlations. However our forecasts may nevertheless be optimistic because we adopted a purely statistical analysis without accounting for any systematic errors. A focus of current \cite{Dalang:2023ehp,Leyde:2024tov,Dalang:2024gfk} and future work should be to study robustness of the cross-correlation method to systematic biases.

\begin{acknowledgments}
JP, DH, and DJ were supported by the NSF under contract AST-2307026. JP, DH and CA also acknowledge support from the Leinweber Center for Theoretical Physics and DOE under contract DE-SC009193. JP thanks Otávio Alves for many valuable conversations. This research was supported in part through computational resources and services provided by Advanced Research Computing at the University of Michigan, Ann Arbor. 
\end{acknowledgments}

\appendix

\section{spectral siren method degeneracy}
\label{app:spectral_siren}

\begin{table}[t]
\centering
\begin{tabular}{c c l}
\hline
$N$ & $\sigma(H_0)/H_0$ & Comment \\
\hline
0   & 2.4\%  & Baseline, no evolution \\
%1--3 & 2.1\% & Finite, mildly degraded \\
4+  & $\infty$ & Fisher singular (degenerate) \\
\hline
\end{tabular}
\caption{Illustrative spectral siren constraints for $N_{\rm GW}=1000$. Adding redshift–evolution terms inflates $\sigma(H_0)$, since their effects increasingly mimic the impact of changing $H_0$. Once the two become indistinguishable, the constraint on $H_0$ diverges.}
\label{tab:spectral_siren}
\end{table}
In order to further motivate the focus of this paper on the dark GW sirens, we briefly illustrate potential challenges with another recently discussed method - that of spectral sirens \cite{Chernoff:1993th,Taylor:2012db,Farr:2019twy,You:2020wju,Mastrogiovanni:2021wsd,Ezquiaga:2022zkx,Mali:2024wpq}. In this approach, one considers the mass spectrum of GW events and, assuming it evolves in time in a way that is known or can be modeled, uses its temporal evolution to place constraints on the Hubble constant or other cosmological parameters. 

We test $H_0$ inference from spectral siren method allowing for the possibility that source-frame mass spectrum features evolve with redshift. Previous studies have shown that such evolution can bias cosmological inference \cite{Agarwal:2024hld,Tong:2025xvd}. Here, we show that redshift evolution of source-frame mass spectrum is degenerate with $H_0$, leading to inflated uncertainties or, in the fully degenerate limit, a complete loss of $H_0$ information. 

We simulate a GW catalog adopting the \textsc{Power Law + Peak} model \citep{Talbot:2018cva} 
as our fiducial case and bin events in observed luminosity distance. We note that source–frame masses are defined by $m_{\rm src}=m_{\rm det}/(1+z)$, where $m_{\rm det}$ is the detector–frame mass. Varying $H_0$ repopulates $d_L$ bins and reshapes the mass–spectrum counts within each distance bin through $m_{\rm det}$.

To study effects of possible redshift evolution, we consider the mean (of the Gaussian peak) of source–frame mass in the spectrum -- that is, the number density vs.\ mass relation, $N^{\mathrm{mass}}(M)$. We model this mean mass by 
\begin{equation}
    \mu(z) \;=\; \mu_0 \;+\; \sum_i c_i \,(1+z)^{\gamma_i},
\end{equation}
where $\mu_0$ is the fiducial Gaussian mean, and 
the coefficients $c_i$ with exponents $\gamma_i$ parameterize redshift evolution of the 
spectral feature. The Fisher matrix for the observations of the mass spectrum (that is number counts vs mass) is

\begin{equation}
    F_{\alpha\beta}
    = \sum_i 
    \frac{1}{\sigma_{N,i}^2}
    \frac{\partial N_i^{\mathrm{mass}}}{\partial \theta_\alpha}
    \frac{\partial N_i^{\mathrm{mass}}}{\partial \theta_\beta},
\end{equation}
where $i$ runs over the bins in mass. Here $\theta_\alpha \in \{H_0, c_1, c_2, \dots\}$ are parameters of the model including the Hubble constant, and $\sigma_{N,i}^2$ is the variance of the mass-spectrum counts (for Poisson noise). If the derivatives with respect to the nuisance parameters,
$\{\partial \boldsymbol{N}^{\mathrm{mass}} / \partial c_i\}$,
can be linearly combined to reproduce the derivative with respect to $H_0$,
then the Fisher matrix will become singular and the forecasted uncertainty on $H_0$ will diverge.

To illustrate this with some results, we perform a simple test case assuming $N_{\rm GW}=1000$ 
dark sirens (see \cref{tab:spectral_siren}). 
We compare a baseline scenario with no redshift evolution
to cases in which additional $(1+z)^{\gamma}$ terms are included. 
As more evolution terms are added, their response vectors can be combined to approximate the $H_0$ response with increasing accuracy. Once this alignment becomes exact, the Fisher matrix becomes singular and the uncertainty in $H_0$ diverges. In our setup, we used eight luminosity distance bins and 24 mass spectrum bins. We found a four-term basis, i.e., $c_i(1+z)^{\gamma_i}$ for $i=1,\dots,4$, renders the 
Fisher matrix numerically singular, and the $H_0$ uncertainty diverges.

This toy example shows that the spectral siren method can be highly sensitive 
to assumptions about the redshift evolution of source-frame spectral features. Even modest, smooth redshift evolution can eliminate the $H_0$ information entirely.

\section{$H_0$ from cross-correlations}
\label{app:H0_constraint}

In the computation of the cross-correlation function \( C^{(\rm gal,gw)}_{ij}(\ell) \), the functions \( \alpha^{\rm gal}_i (\ell, k, H_0) \) play a crucial role, where \( x \) and \( y \) denote different data sets—galaxies and GW events, respectively. For galaxies, redshifts \( z \) are measured directly and converted into comoving distances \( r \) using the fiducial cosmology. When expressed in units of \( h^{-1}\,\mathrm{Mpc} \) (where \( h = H_0 / (100\,\mathrm{km\,s^{-1}\,\mathrm{Mpc}^{-1}}) \)), the dependence on the Hubble constant \( H_0 \) cancels out. This renders the galaxy distribution \( n^{\rm gal}_i(r) \) independent of \( H_0 \):
\begin{equation}
n^{\rm gal}_i(r) \propto \frac{dN^{\rm gal}}{dr},
\end{equation}

where \( dN^{\rm gal}/dr \) is defined in Eq.~\eqref{eq:dN_dr_gal}. The \( \alpha \) function for galaxies is independent of $H_0$ and given by:

\begin{equation}
\alpha^{\rm gal}_i (\ell, k) = b_i^{\rm gal} \int_0^\infty dr \, n^{\rm gal}_i(r) \, j_\ell(kr),
\end{equation}
where \( b^{\rm gal} \) is the galaxy bias and \( j_\ell(kr) \) is the spherical Bessel function of order \( \ell \).

For GW events, distances are measured via luminosity distances \( D_L \), which are related to comoving distances by:

\begin{equation}
D_L = (1 + z) \, r.
\end{equation}

Since the distance of GW events are observed directly in unit of Mpc, converting \( D_L \) to \( r \) in unit of $h^{-1}$Mpc necessitates an assumption about \( H_0 \). This introduces an explicit dependence of the GW source distribution \( n^{\rm gw}_j(r, H_0) \) on \( H_0 \):

\begin{equation}
n^{\rm gw}_j(r, H_0) \propto \frac{dN^{\rm gw}_j}{dr},
\end{equation}
where \( N^{\rm gw}_j \) is the number of GW events in the \( j \)-th distance bin. The \( \alpha \) function for GW events then becomes

\begin{equation}
\alpha^{\rm gw}_j (\ell, k, H_0) = b_j^{\rm gw} \int_0^\infty dr \, n^{\rm gw}_j(r, H_0) \, j_\ell(kr),
\end{equation}
with \( b_j^{\rm gw} \) being the GW bias parameter.

Substituting these into the cross-correlation function:
\begin{widetext}
\begin{equation}
C^{\rm gal,gw}_{ij}(\ell) = 4\pi \int_0^\infty \frac{dk}{k} \, \alpha^{\rm gal}_i (\ell, k) \, \alpha^{\rm gw}_j (\ell, k, H_0) \, \Delta_{ij}^2(k)+  P_{\ell,\mathrm{shot}}^{\,\mathrm{gal}_i\mathrm{gw}_j}.
\end{equation}
\end{widetext}

Note that we fix $\Delta_{ij}^2(k)$ to a fixed cosmology, so it does not depend on $H_0$. We see that \( C^{\rm gal,gw}_{ij}(\ell) \) depends on \( H_0 \) through \( \alpha^{\rm gw}_j (\ell, k, H_0) \). The galaxy term \( \alpha^{\rm gal}_i (\ell, k) \) is fixed and independent of \( H_0 \) in \( h^{-1}\,\mathrm{Mpc} \) units, whereas the GW term retains \( H_0 \) dependence due to the conversion from luminosity distance to comoving distance.

Only the correct value of \( H_0 \) will properly align the GW events with the galaxy distribution in comoving space, maximizing the true cross-correlation signal. By analyzing how \( C^{\rm gal,gw}_{ij}(\ell) \) varies with different assumed values of \( H_0 \), we can constrain the true value of the Hubble constant through observational data.

\section{Shot Noise with Radial and Angular Smoothing}
\label{app:shot_noise_smoothed}

In our analysis both galaxy (\(X=\mathrm{gal}\)) and GW (\(X=\mathrm{gw}\)) samples are smoothed  
radially by Gaussian kernels \(W^X_i(r)\) (Eq.~\ref{eq:window}) and have finite angular localization uncertainty \(W^X_\ell\) (Eq.~\ref{eq:angular_window}). We define the \emph{mean} number density per steradian in the \(i\)th radial bin as
\begin{equation}
\bar N^{\Omega}_{X,i}
\;\equiv\;
\frac{\mathrm{d}N^X_i}{\mathrm{d}\Omega}
\;=\;
\frac{1}{\Omega_{\rm survey}}
\int_{0}^{\infty}
n^X(r)\;W^X_i(r)\,
\mathrm{d}r,
\label{eq:dNi_dOmega_bar}
\end{equation}
where \(n^X(r)\) is the un‐binned comoving‐distance density (Eqs.~\ref{eq:galaxy_n}, \ref{eq:gw_n}) and \(\Omega_{\rm survey}\) is the total survey solid angle.

\subsection{Auto–correlation for Galaxies and GW}

Let the discrete projected count in direction \(\hat n\) be
\begin{equation}
\mathcal{N}^X_i(\hat n)
\;\equiv\;\frac{1}{\Omega_{\rm survey}}\int d^2 \hat{n}_k B(\hat{n}_k)\sum_{k\in i}\delta^{(2)}\!\bigl(\hat n-\hat n_k\bigr)\,,
\label{eq:proj_count}
\end{equation}

We then form the overdensity field
\begin{equation}
\delta^X_i(\hat n)
\;=\;
\frac{\mathcal{N}^X_i(\hat n)\;-\;\bar N^{\Omega}_{X,i}}
     {\bar N^{\Omega}_{X,i}}
\label{eq:overdensity}
\end{equation}

Expanding in spherical harmonics,
\begin{equation}
\delta^X_i(\hat n)
= \sum_{\ell=0}^\infty\sum_{m=-\ell}^{\ell}
  a^{X_i}_{\ell m}\;Y_{\ell m}(\hat n),
\end{equation}
Note that the angular beam \(W^X_\ell\) do not convolve with the auto shot noise unlike the case for cross terms of shot noise.
Including \(W^X_\ell\) in the auto shot noise would artificially remove the impact of angular localization uncertainty, since the beam suppression would cancel out in the Fisher matrix.
% \begin{equation}
% a_{\ell m}^{X_i,\mathrm{obs}}
% =
% W^X_\ell\;\int d\Omega
% \delta^X_i(\hat{n}) Y^*_{\ell m}(\hat n_k).
% \end{equation}

% Assuming Poisson sampling, the shot‐noise correlator is
% \begin{equation}
% \bigl\langle
% a_{\ell m}^{X_i,\mathrm{obs}}\,
% a_{\ell' m'}^{X_i,\mathrm{obs}*}
% \bigr\rangle_{\rm shot}
% =
% (W^X_\ell)^2\,
% \frac{1}{\bar N^{\Omega}_{X,i}}
% \;\delta_{\ell\ell'}\,\delta_{m m'}.
% \end{equation}
Thus, the auto–correlation shot‐noise power is
\begin{equation}
P_{\ell, \rm shot}^{\,X_iX_i}
\;=\;
\frac{1}{\bar N^{\Omega}_{X,i}}\,.
\end{equation}

\subsection{Cross–correlation between galaxies and GW}
For a galaxy bin \(i\) and a GW bin \(j\), only the objects common to both contribute to the disconnected (shot‐noise) term.  First, we define their \emph{radial} overlap density,
\begin{equation}
n^{\rm both}_{ij}(r)
\;\equiv\;
\min\!\bigl[n^{\rm gal}(r)\,W^{\rm gal}_i(r)\,,\;n^{\rm gw}(r)\,W^{\rm gw}_j(r)\bigr].
\label{eq:nboth_min}
\end{equation}

Next introduce the \emph{radial} overlap mean density (per steradian),
\begin{equation}
\bar N^{\Omega,\,\rm both}_{ij}
\;\equiv\;
\frac{\mathrm{d}N^{\rm both}_{ij}}{\mathrm{d}\Omega}
\;=\;
\frac{1}{\Omega_{\rm survey}}
\int_{0}^{\infty}
n^{\rm both}_{ij}(r)\,
\mathrm{d}r.
\label{eq:Nboth}
\end{equation}

Carrying through the same projection and beam‐smoothing steps as in the auto‐correlation, the cross–correlation shot‐noise term is then
\begin{equation}
P_{\ell, \rm shot}^{\mathrm{gal}_i\,\mathrm{gw}_j}
=
W^{\mathrm{gal}}_\ell\;W^{\mathrm{gw}}_\ell
\;
\frac{\bar N^{\Omega,\,\rm both}_{ij}}
     {\bar N^{\Omega}_{\mathrm{gal},i}\,\bar N^{\Omega}_{\mathrm{gw},j}}.
\end{equation}

If the two radial bins do not overlap, then \(N^{\rm both}_{ij}(r)\equiv0\) and hence $P_{\ell,\rm shot}^{\mathrm{gal}_i,\mathrm{gw}_j}=0$.  

For galaxy samples, the conversion from redshift to comoving distance is performed in units of $h^{-1}\mathrm{Mpc}$, so there is no dependence on $H_0$ in either $n^{\mathrm{gal}}(r)$ or $\bar N^{\Omega}_{\mathrm{gal},i}$. For GW samples, although the conversion from luminosity distance to comoving distance introduces an explicit $H_0$ dependence in $n^{\mathrm{gw}}(r)$, this dependence cancels out in $\bar N^{\Omega}_{\mathrm{gw},i}$ after integrating over $r$. Thus, the projected number density per steradian $\bar N^{\Omega}_{X,i}$ is independent of $H_0$ for both tracers.

However, for the cross-correlation shot-noise term, the situation is different. In Eq.~\eqref{eq:Nboth}, the numerator $\bar N^{\Omega,\,\mathrm{both}}_{ij}$, which is the integrated overlap of the galaxy and GW distributions, retains dependence on $H_0$. This is because the overlap distribution $\min[n^{\rm gal}(r)\,W^{\rm gal}_i(r),\, n^{\rm gw}(r)\,W^{\rm gw}_j(r)]$ depends on the relative radial alignment of the two tracers, and the conversion from GW luminosity distance to comoving distance is $H_0$-dependent, which cannot be canceled out.

Therefore, only the cross-correlation shot-noise term $P_{\ell, \rm shot}^{\mathrm{gal}_i\,\mathrm{gw}_j}$ carries information about $H_0$, enabling us to constrain $H_0$ from the shot-noise contribution in cross-correlation, while the auto-correlation shot noise terms remain $H_0$-independent.

\section{Radial smoothing}
\label{app:radial_smoothing}

When a field on the sphere is smoothed (or convolved) with a Gaussian beam, the operation in real (angular) space becomes a simple multiplication in spherical harmonic space. This section illustrates the derivation of the angular window function and shows how the convolution translates to a product in harmonic space, with an application to the final expression for the cross-correlation.

The convolution theorem states that a convolution in real space becomes a product in Fourier (or harmonic) space. For example, if a function \(f(\mathbf{x})\) is convolved with a kernel \(g(\mathbf{x})\),
\begin{equation}
(f \ast g)(\mathbf{x}) = \int \, g(\mathbf{x}-\mathbf{x}')\, f(\mathbf{x}')d^3\mathbf{x}',
\end{equation}
then its Fourier transform is
\begin{equation}
\widetilde{(f \ast g)}(\mathbf{k}) = \tilde{g}(\mathbf{k})\, \tilde{f}(\mathbf{k}).
\end{equation}

Here we demonstrate that the normalized number density distribution 
\begin{equation}
    n_i(r) = \frac{1}{N^{\rm gal}}\frac{dN^{\rm gal}}{dr}\,W_i(r)
    \label{eq:nr_definition}
\end{equation}
(which is obtained by summing over the contributions of all events, each of them is a Gaussian distribution) is equivalent to a distribution that has been smoothed by a radial Gaussian kernel, which is defined as the following.
\begin{equation}
S_r(r,r') = \frac{1}{\sqrt{2\pi}\sigma_r}\,\exp\left[-\frac{(r-r')^2}{2\sigma_r^2}\right],
\end{equation}

Given the smoothing kernel, the effective (smoothed) distribution is
\begin{equation}
    n_{\rm eff}(r) = \int dr'\; S_r(r, r')\, n_{\rm true}(r'),
    \label{eq:convolution}
\end{equation}
where \( n_{\rm true}(r) \) represents the underlying true distribution of sources. 

The true radial distribution of events is described by a sum of Dirac delta functions (equivalent to infinite precision in localization),
\begin{equation}
    n_{\rm true}(r) = \sum_i \delta(r - r_i),
    \label{eq:true_distribution}
\end{equation}
where \( r_i \) is the true radial coordinate of the \(i\)th event.

Due to observational uncertainties, the measured radial position of each event is not exact but rather is distributed around \(r_i\) according to a Gaussian probability density,
\begin{equation}
    p_i(r) = \frac{1}{\sqrt{2\pi}\sigma_r}\,\exp\left[-\frac{(r - r_i)^2}{2\sigma_r^2}\right] = S_r(r, r_i).
    \label{eq:event_distribution}
\end{equation}
Thus, the contribution of each event to the observed number density is exactly the kernel \(S_r(r, r_i)\).

Summing over all events, the observed (or effective) number density distribution is
\begin{widetext}
\begin{align}
    \frac{dN^{\rm gal}}{dr} &= \sum_i p_i(r) \nonumber\\[1mm]
         &= \sum_i S_r(r, r_i) \nonumber\\[1mm]
         &= \sum_i \int dr'\; S_r(r, r')\, \delta(r' - r_i) \\[1mm]
         &= \int dr'\; S_r(r, r')\, \left[\sum_i \delta(r' - r_i)\right] \nonumber\\[1mm]
         &= \int dr'\; S_r(r, r')\, n_{\rm true}(r'),\nonumber
    \label{eq:observed_vs_convolved}
\end{align}
\end{widetext}
which is exactly the convolution expression in Eq.~\eqref{eq:convolution}.

The observed number density distribution is also normalized and include window function \(W_i(r)\) that encodes binning as in Eq.~\eqref{eq:nr_definition}. However, these two factors don't influence the derivation here. Since each individual event’s uncertainty has already been incorporated through its Gaussian profile \(S_r(r, r_i)\), the overall distribution \(n(r)\) is effectively a sum of these smoothed contributions. Hence, the radial smoothing is already built into the definition of \(n(r)\) 

\begin{equation}
n_i(r) =  \frac{1}{N^{\rm gal}}\int dr'\; S_r(r, r')\, n_{\rm true}(r')\,W_i(r).
\end{equation}
Therefore, when radial uncertainties are already incorporated into the individual contributions \(p_i(r)\), no additional smoothing is necessary.

\section{Angular power spectrum including angular smoothing}  
\label{app:angular_smoothing}

Consider a beam (smoothing kernel) on the sphere with a Gaussian profile,
\begin{equation}
B(\theta) = \frac{1}{2\pi \sigma_\theta^2}\exp\left(-\frac{\theta^2}{2\sigma_\theta^2}\right),
\end{equation}
where \(\theta\) is the angular separation and \(\sigma_\theta\) is the beam width (in radians). Its spherical harmonic coefficients are given by
\begin{equation}
B_\ell = 2\pi \int_0^\pi d\theta\, \sin\theta\, B(\theta)\, P_\ell(\cos\theta),
\end{equation}
with \(P_\ell(\cos\theta)\) being the Legendre polynomials. 

For a narrow beam (small \(\sigma_\theta\)) the dominant contribution comes from small \(\theta\), where one may approximate \(\sin\theta \approx \theta\) and \(\cos\theta \approx 1 - \theta^2/2\). 
Under these approximations (or by referring to standard results in CMB analyses \cite{White:1994ru}), one obtains
\begin{equation}
B_\ell \approx \exp\left[-\frac{1}{2}\ell(\ell+1)\sigma_\theta^2\right].
\end{equation}
This factor, commonly denoted as \(W_\ell\), is the angular window function:
\begin{equation}
W_\ell = \exp\left[-\frac{1}{2}\ell(\ell+1)\sigma_\theta^2\right].
\label{eq:angular_window}
\end{equation}

The physical meaning is as follows: if a sky map \(f(\hat{n})\) (with \(\hat{n}\) a direction on the sphere) is convolved with the beam \(B(\theta)\), its spherical harmonic coefficients \(a_{\ell m}\) transform as
\begin{equation}
a_{\ell m}^{\rm convolved} = B_\ell\, a_{\ell m}^{\rm original} = W_\ell\, a_{\ell m}^{\rm original}.
\end{equation}

In our analysis, the \(\alpha\) function with unsmoothed angular function for GW events is defined as
\begin{equation}
\alpha^{\rm gw}_j (\ell, k, H_0) = b_j^{\rm gw} \int_0^\infty dr\, n^{\rm gw}_j(r, H_0)\, j_\ell(kr).
\end{equation}
Including the angular smoothing (and note that the radial uncertainties are already incorporated in \(n^{\rm gw}_j(r, H_0)\)), we modify the \(\alpha\) function by multiplying by \(W_\ell\):
\begin{equation}
\alpha^{\rm gw}_{j} (\ell, k, H_0) = b_j^{\rm gw}\,W^{\rm gw}_\ell \int_0^\infty dr\, n^{\rm gw}_j(r, H_0)\, j_\ell(kr).
\end{equation}
One would obtain similar results for the galaxies.

Substituting this into the cross-correlation function,
\begin{equation}
P^{(\rm gal,gw)}_{ij}(\ell) = 4\pi \int_0^\infty \frac{dk}{k} \, \alpha^{\rm gal}_i (\ell, k) \, \alpha^{\rm gw}_j (\ell, k, H_0) \, \Delta_{ij}^2(k),
\end{equation}
and the final expression including the angular smoothing for the galaxy-GW cross-correlation becomes
\begin{widetext}
\begin{equation}
C^{(\rm gal,gw)}_{ij}(\ell) = 4\pi\, b_i^{\rm gal}\, b_j^{\rm gw}\, W^{\rm gal}_\ell W^{\rm gw}_\ell \int_0^\infty \frac{dk}{k} \left[ \int_0^\infty dr\, n^{\rm gal}_i(r)\, j_\ell(kr) \right] \left[ \int_0^\infty dr\, n^{\rm gw}_j(r, H_0)\, j_\ell(kr) \right] \Delta_{ij}^2(k)+  P_{\ell,\mathrm{shot}}^{\,\mathrm{gal}_i\mathrm{gw}_j}.
\end{equation}
\end{widetext}

In this final expression, the factor \(W_\ell = \exp\left[-\tfrac{1}{2}\ell(\ell+1)\sigma_\theta^2\right]\) accounts for the angular smoothing (i.e., the finite resolution in the sky positions).

 We assume a galaxy survey with 1 arcsecond resolution. One arcsecond is
\[
1\,\mathrm{arcsec} \approx \frac{1}{3600}\,\mathrm{deg} \approx 4.85\times10^{-6}\,\mathrm{rad}.
\]
Thus, a representative value for $\sigma_\theta$ in a galaxy survey might be
\[
\sigma_\theta \sim 5\times10^{-6}\,\mathrm{rad}.
\]

In contrast, current ground‐based GW detectors (LIGO--Virgo) typically localize events with error regions of tens to hundreds of square degrees \cite{KAGRA:2013rdx}. If one approximates a GW event's localization as a Gaussian probability distribution over a flat sky, the one–sigma angular resolution is given by
\[
\sigma_\theta \sim \sqrt{\frac{A}{\pi}}.
\]
In our fiducial case, we assume an GW event localized to roughly $A\approx 10\,\mathrm{deg}^2$, and
we have
\[
\sigma_\theta \sim \sqrt{\frac{10\times (0.01745)^2}{\pi}} \approx 0.03\,\mathrm{rad}\quad.
\]

\section{Effects of galaxy redshift distribution}
\label{app:redshift_Dis}

\begin{figure*}[htbp]
  \centering
  \newlength{\imgsep}\setlength{\imgsep}{0.008\textwidth} % gap between panels
  \begin{minipage}[b]{\dimexpr(\textwidth-2\imgsep)/3\relax}
    \centering
    \includegraphics[width=\linewidth,trim=8pt 8pt 8pt 8pt,clip]{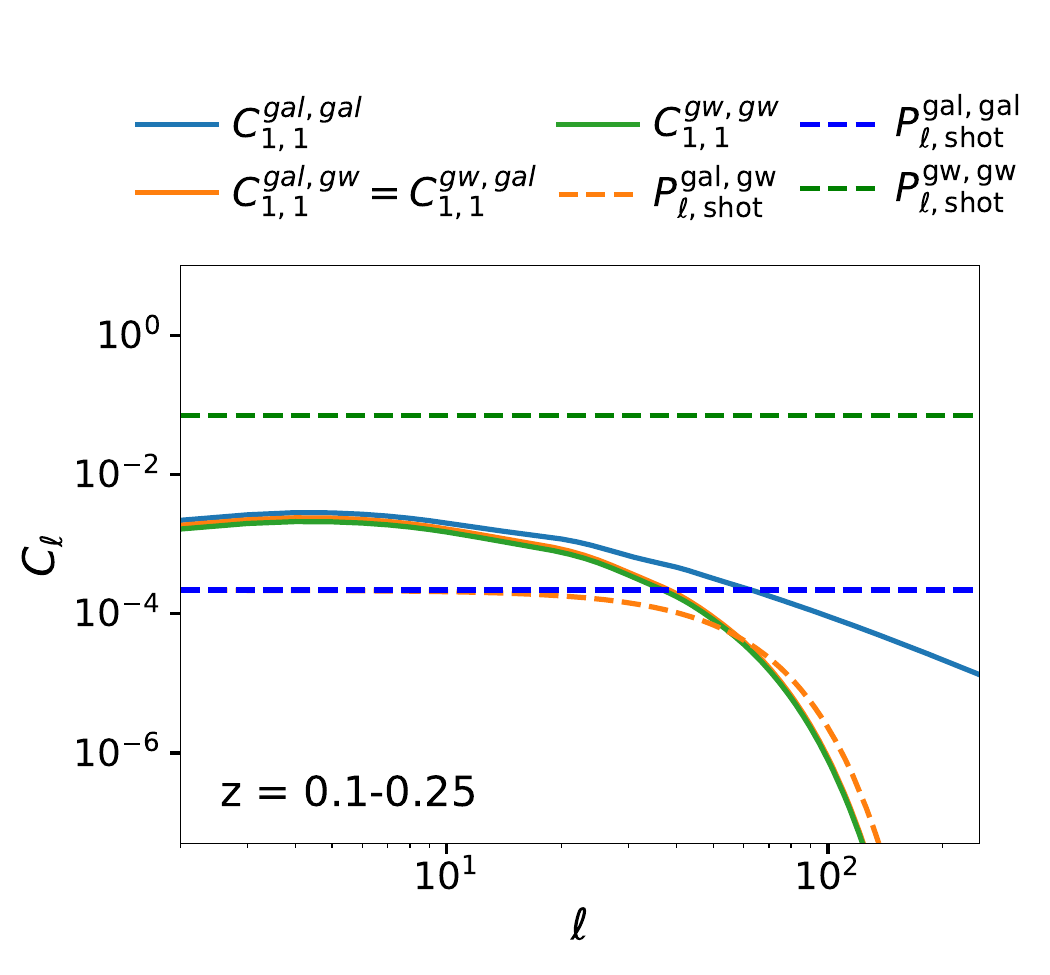}
    \label{fig:redshift_low}
  \end{minipage}\hspace{\imgsep}%
  \begin{minipage}[b]{\dimexpr(\textwidth-2\imgsep)/3\relax}
    \centering
    \includegraphics[width=\linewidth,trim=8pt 8pt 8pt 8pt,clip]{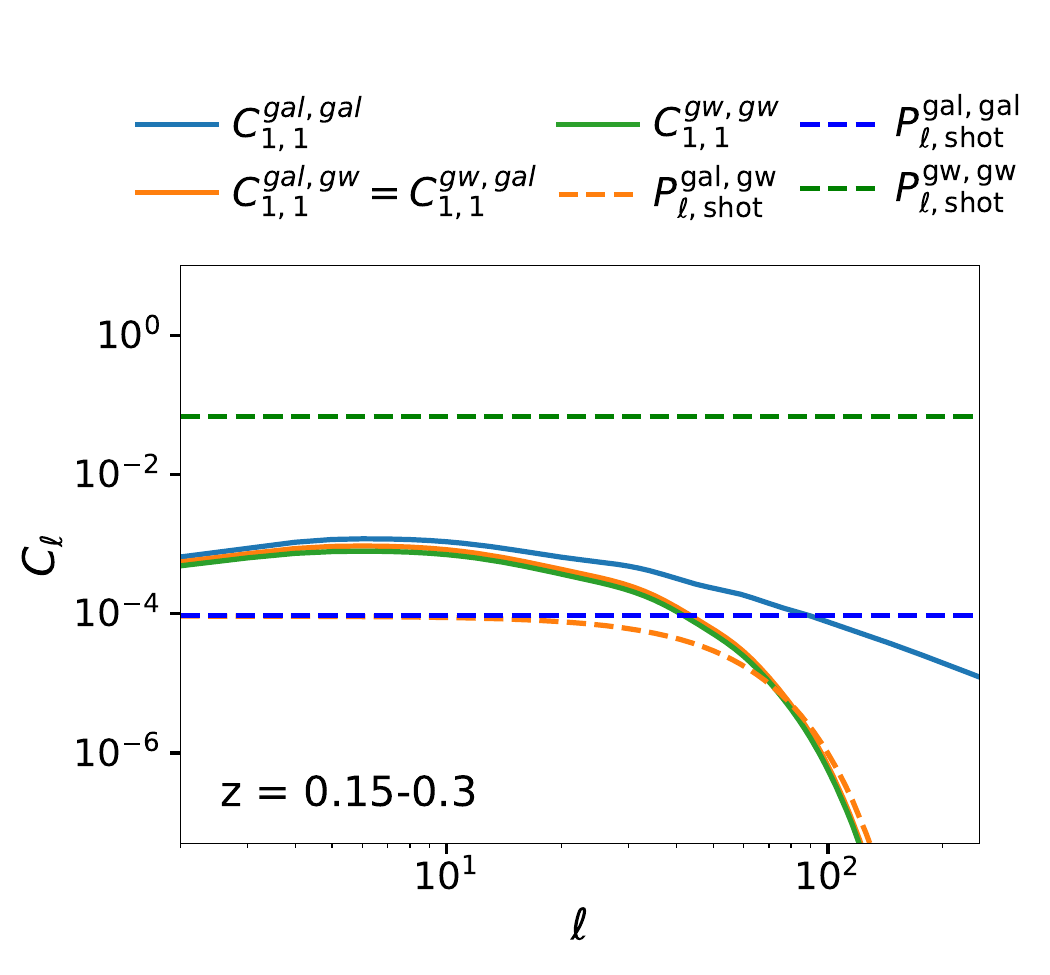}
    \label{fig:redshift_mid}
  \end{minipage}\hspace{\imgsep}%
  \begin{minipage}[b]{\dimexpr(\textwidth-2\imgsep)/3\relax}
    \centering
    \includegraphics[width=\linewidth,trim=8pt 8pt 8pt 8pt,clip]{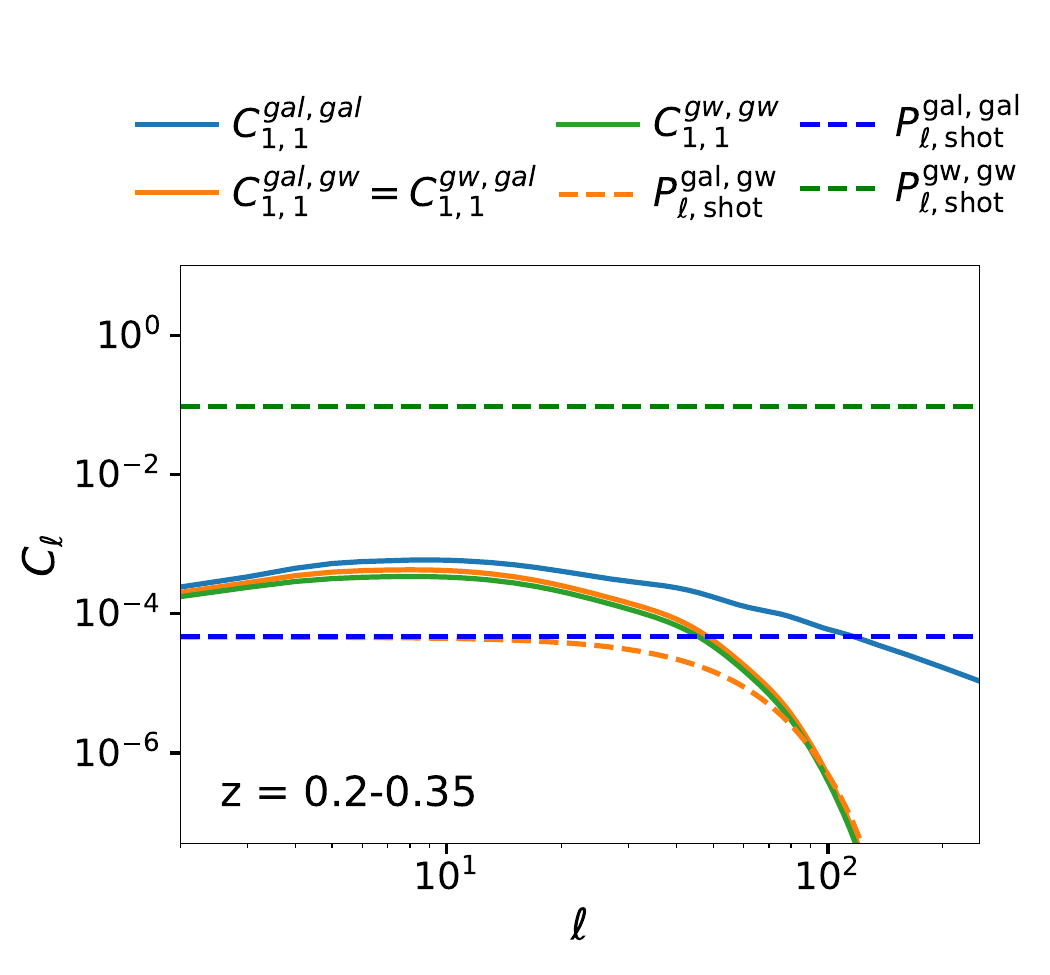}
    \label{fig:redshift_high}
  \end{minipage}
  \caption{Angular power spectra in three different redshift ranges assuming the fiducial cosmology in the first bin with a total of 15 bins.}
  \label{fig:redshift_effects}
\end{figure*}

Figure~\ref{fig:redshift_effects} compares the angular power spectra and shot-noise terms for three redshift selections assuming 15 bins, $N_{\rm GW}=1{,}000$, a fixed galaxy number density ($5\times10^{-4}\,h^{3}\,\mathrm{Mpc}^{-3}$), and $\ell_{\max}=250$: 0.1–0.25, 0.15–0.30, and 0.20–0.35. In this test, GW events are sampled from the same galaxy redshift distribution. We find that $H_0$ constraints are significantly tighter at low redshift. This is because the luminosity distance errors are smaller and the cross-correlation signal is stronger, which together enhance the constraining power. At higher redshift, although the survey volume and number of galaxies increase and the shot-noise contribution decreases slightly, the clustering signal weakens substantially. Since the cross-correlation carries the dominant information for $H_0$, this weakening of the clustering signal leads to degraded constraints overall. Thus, extending to higher redshift without improving distance precision or event statistics does not improve $H_0$ constraints, and low-$z$ GW samples remain especially valuable.

\section{Effects of GW distribution}
\label{app:GW_Dis}

% \jiaming{Damon Introduce the GWTC-3 GW distributions calculation}
In our main forecasts we assume GW events are sampled from the galaxy redshift distribution.  To test a more realistic scenario, we instead draw events from the GWTC-3–inferred GW distribution and compute each event’s $\sigma_{d_L}$\footnote{In this test, we assume constant sky localization error. Alternatively, one can calculate each event's sky localization error based on its SNR.} (details of the calculation are provided below). 

To simulate the GW population, we follow the population model described in~\citep{KAGRA:2021duu}. 
We assume the \textsc{PowerLaw+Peak} model for the primary ($m_1$) and secondary ($m_2$) mass distributions. 
\begin{equation}
    p_{\mathrm{GW}}(z) \propto \frac{1}{1+z} \frac{dV_c}{dz} (1+z)^\kappa,
\end{equation}
where $\kappa$ is the governing parameter of the distribution.
All parameter values are chosen to be the best-fit parameters from~\citep{KAGRA:2021duu}. 
After simulating the intrinsic $(m_1, m_2, z)$ population, we apply the detectability $p_\mathrm{det}(m_1, m_2, z)$ to obtain the observed population. 
For simplicity, we use the package \textbf{gwdet}~\citep{davide_gerosa_2017_889966}. 
Finally, we convert the redshift to luminosity distance using the fiducial model.

\begin{figure}[tbp]
  \centering
  \includegraphics[width=\linewidth]{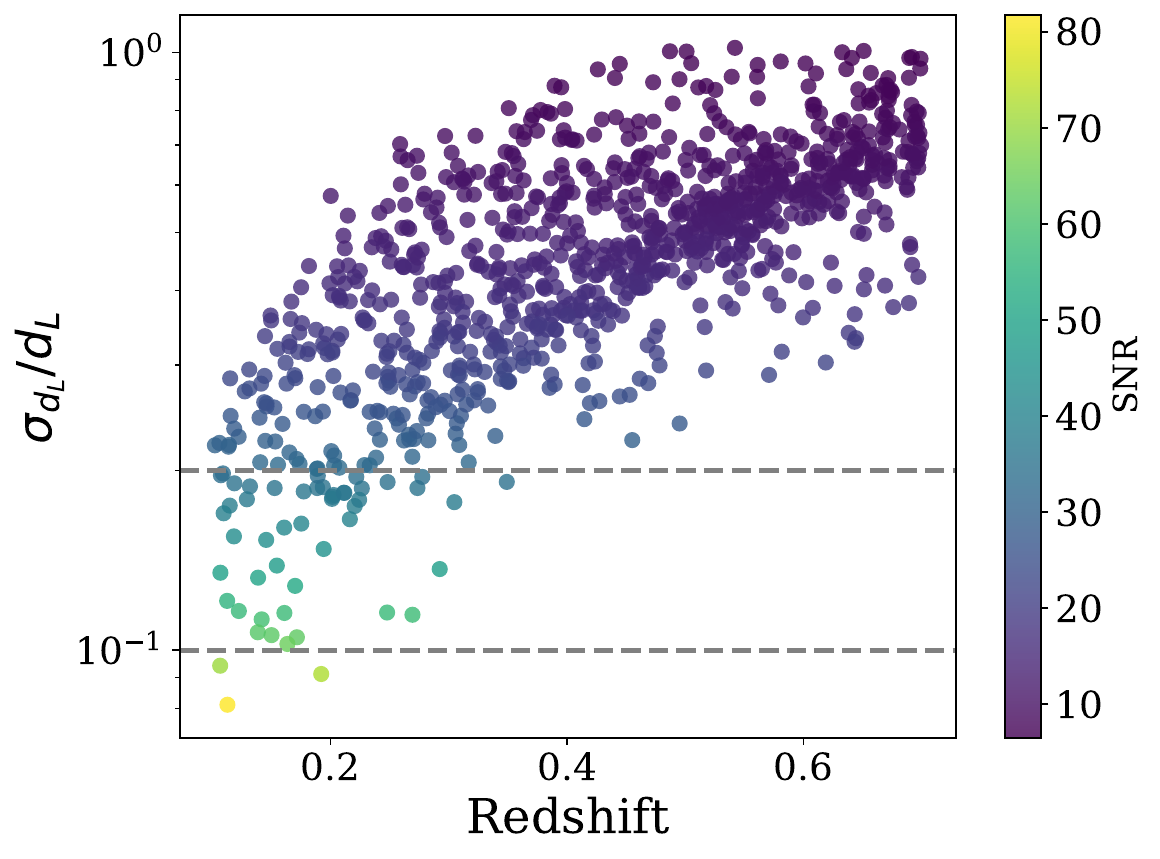}
  \caption{Relative uncertainty $\sigma_{d_L}/d_L$ versus redshift $z$ for our GWTC-3-inferred GW sample. Each point is color-coded by its SNR, as shown by the accompanying color bar. The dashed horizontal lines, with coordinate values $\sigma_{d_L}/d_L = 0.1$ and $0.2$, indicate the two distance precisions adopted in this test.}
  \label{fig:sigma_dL_vs_z_gw}
\end{figure}

For each GW event, we simulate the luminosity distance ($d_\mathrm{L}$) posterior using a Gaussian distribution with a standard deviation $\sigma_{d_\mathrm{L}}(d_\mathrm{L})$, which is inversely proportional to the SNR.
The relationship between $\sigma_{d_\mathrm{L}}(d_\mathrm{L})$ and SNR is fitted using events from GWTC-3~\citep{KAGRA:2021duu}. The simulated relative luminosity distance uncertainties and corresponding SNR values are shown in~\cref{fig:sigma_dL_vs_z_gw}.

In our model, the $d_\mathrm{L}$ measurement uncertainty depends on $(m_1, m_2, d_\mathrm{L}(z))$. 
To incorporate this into the window function, we marginalize over $m_1$ and $m_2$, weighted by their population distribution:

\begin{equation}
    \sigma_{d_\mathrm{L}}(d_\mathrm{L}) = \int \sigma(m_1,m_2|d_\mathrm{L}) p_{\mathrm{GW}}(m_1, m_2) dm_1 dm_2.
\end{equation}

\begin{figure*}[t]
\includegraphics[width=\linewidth]{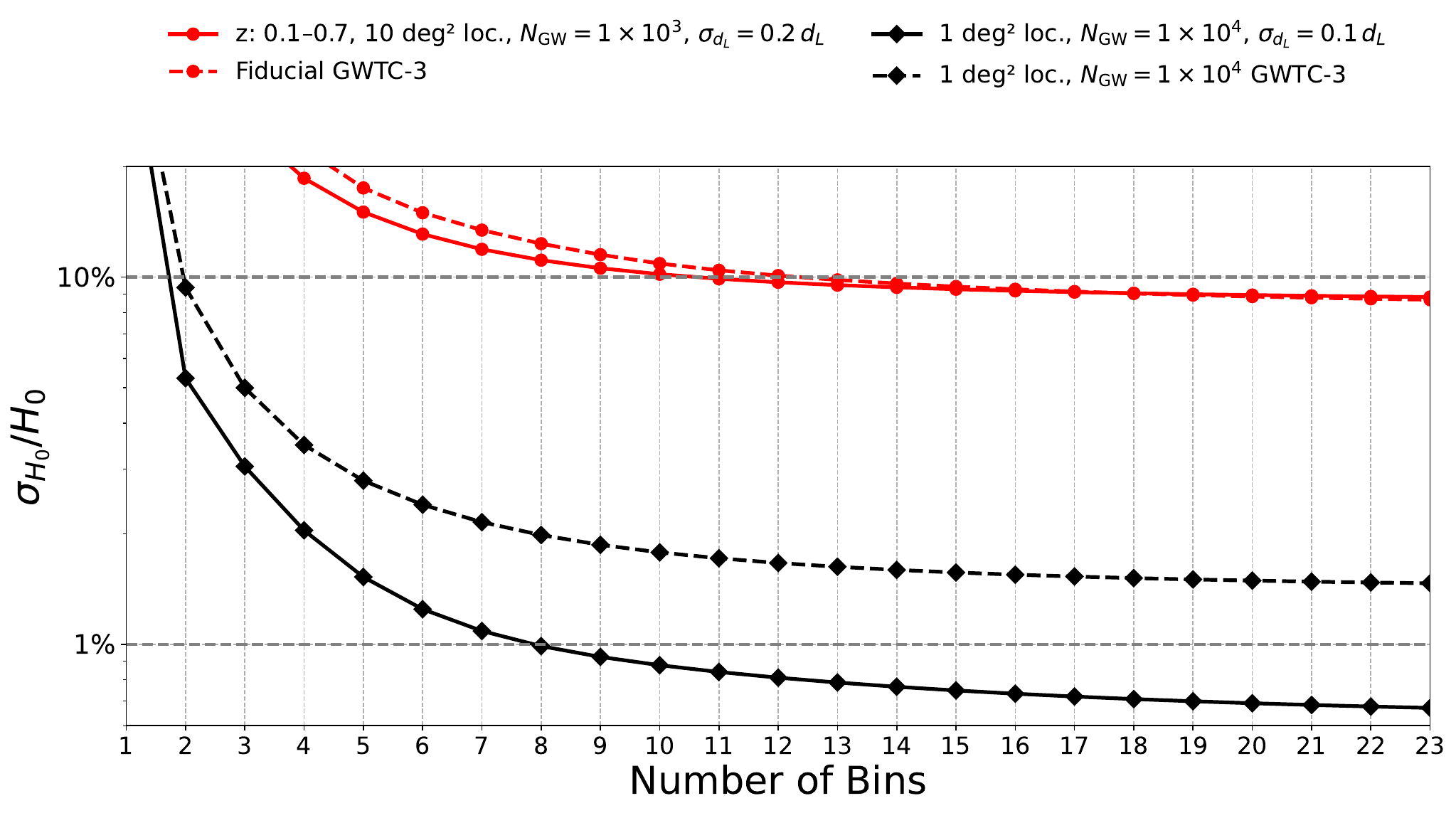}
\caption{ Relative uncertainty in $H_0$ for two scenarios: (i) fiducial localization ($10\, \mathrm{deg}^2$, $N_{\rm GW}=1000$, solid red) and (ii) idealized case ($1\, \mathrm{deg}^2$, $N_{\rm GW}=10,000$, solid black). Solid curves correspond to GW events drawn from the galaxy redshift distribution; dashed curves correspond to sampling from the GWTC-3 redshift distribution with each event’s SNR setting $\sigma_{d_L}$.
}
\label{fig:H0_LVK}
\end{figure*}

In \cref{fig:H0_LVK}, we compare $H_0$ constraints obtained from GW events sampled from the galaxy redshift distribution with those from the GWTC-3 redshift distribution, where each event’s $\sigma_{d_L}$ is set by its SNR. Compared to our fiducial model with $\sigma_{d_L}=0.2d_L$, GWTC-3–inferred distribution has more distant GW events with larger distance uncertainties, and about $5\%$ of events have $\sigma_{d_L}<0.2d_L$ at low redshift (see~\cref{fig:sigma_dL_vs_z_gw}). Nearly all GWTC-3–inferred GW events have $\sigma_{d_L}$ exceeding $0.1 d_L$. As a result, the $H_0$ constraints from GWTC-3 are identical to those of the fiducial model, but remain weaker than the idealized scenario with $1\,\mathrm{deg}^2$ localization, $\sigma_{d_L}=0.1d_L$, and $N_{\rm GW}=10,000$ (see~\cref{fig:H0_LVK}).

\nocite{*}

\bibliography{main}% Produces the bibliography via BibTeX.

\end{document}